\documentstyle[floats,aps,epsf,amsfonts]{revtex}
\tighten
\begin{document}
\draft


\title{Self-force on a scalar particle in spherically-symmetric spacetime
        via mode-sum regularization: radial trajectories}
\author{Leor Barack}
\address {Department of Physics,
          Technion---Israel Institute of Technology, Haifa, 32000, Israel}
\date{\today}
\maketitle


\begin{abstract}

Recently, we proposed a method for calculating the ``radiation reaction''
self-force exerted on a charged particle moving in a strong field orbit
in a black hole spacetime. In this approach, one first calculates the
contribution to the ``tail'' part of the self force due to each multipole
mode of the particle's self field. A certain analytic procedure is then
applied to regularize the (otherwise divergent) sum over modes. This
involves the derivation of certain regularization parameters using local
analysis of the (retarded) Green's function. In the present paper we
present a detailed formulation of this mode-sum regularization scheme for
a scalar charge on a class of static spherically-symmetric backgrounds
(including, e.g., the Schwarzschild, Reissner-Nordstr\"{o}m, and
Schwarzschild-de Sitter spacetimes). We fully implement the regularization
scheme for an arbitrary radial trajectory (not necessarily geodesic) by
explicitly calculating all necessary regularization parameters in this
case.

\end{abstract}

\section{introduction}\label{secI}

The motion of a test pointlike mass (a ``particle'') in orbit outside a
black hole is commonly studied to model, and gain understanding of,
realistic astrophysical scenarios involving highly relativistic two-body
systems --- particularly, the capture of a small compact object by a
supermassive black hole \cite{Schutz}. To describe the orbital evolution
of such a particle on a strongly curved background, one must take into
account non-geodesic effects caused by the interaction of the particle
with its own gravitational field. This problem of deducing the {\em self
force} (or ``radiation reaction'' force) exerted on the particle is often
treated via perturbation theory: one assumes that the particle is endowed
with a charge $q$ much smaller than the mass of the black hole (this
charge may represent the particle's mass, electric charge or---as in the
toy model studied in the current paper---its scalar charge) and looks for
the $O(q^2)$ correction to the equation of motion. The basic challenging
task involved in this calculation, already in flat space, is, of course,
correctly handling the divergence of the self field at the very location
of the particle; namely, the introduction and justification of an
appropriate regularization method. When considering the case of curved
spacetime, additional difficulty arises due to the {\em nonlocal} nature
of the self-force effect: waves emitted by the particle at some moment may
backscatter off spacetime curvature and interact back with the particle at
later stages of its motion. The occurrence of this so-called ``tail''
contribution to the self force results in that the calculation of the self
force at a given moment requires, in principle, knowledge of the entire
causal history of the particle. A number of methods (briefly surveyed
below) have been proposed over the years for calculating the self force in
curved backgrounds. The interest in this problem has greatly risen lately
by virtue of recent years' developments towards experimental gravitational
waves detection, and the consequent need for accurate predictions for the
orbital evolution of strongly gravitating two-body systems. Yet, actual
calculations of the self force have been restricted thus far only to a few
very simple cases (see below).

The standard technique for calculating the radiative evolution of
orbits around black holes is the one based on energy-momentum balance
considerations \cite{balance}. In this approach one computes the
flux to infinity (and across the horizon) of quantities associated
with the constants of motion in the lack of self-force effects
(specifically, in the Schwarzschild and Kerr backgrounds, the
particle's energy $E$ and azimuthal angular momentum $L_z$), thus
deducing the temporal rate of change of these ``constants''. Such
balance calculations, though developed to a great extent, present two
basic drawbacks: First, in the important Kerr case they are inapplicable
for calculating the rate of change of the third constant of motion
necessary for a full specification of the orbital evolution, i.e., the
Carter constant $Q$, as this quantity is not additive. Second,
these calculations do not account for the non-dissipative, yet
important, part of the self force \cite{Wiseman,Burko3}.

For the above reasons, a method based on direct calculation
of the {\em momentary} force along the worldline seems more adequate.
In the context of electromagnetism in flat space, such a method is
familiar from the classic work by Dirac \cite{Dirac}, concerning the
electromagnetic self force on a (classical) electron.
Dirac avoided the singularity of the self field at the particle's
location by introducing the ``radiative potential'', constructed
by taking the difference between the retarded and advanced electromagnetic
potentials (what results in the cancellation of the problematic singular
part). This procedure gave rise to what is now called the
{\em Abraham-Lorentz-Dirac} (ALD) self force in flat space
[see Eq.\ (\ref{eq2-100}) below for the analogous scalar particle case].
The concept of ``radiative potential'' was much later employed by Gal'tsov
\cite{Gal'tsov} for calculating the temporal rate of change of the energy
and azimuthal angular momentum parameters for electrically charged
particles orbiting a Kerr black hole.
Though Gal'tsov analysis yielded correct results, it seems conceptually
difficult, in general cases, to justify the use of such a non-causal
approach. The problem becomes obvious when considering curved spacetime,
where the self force exhibits nonlocal contributions: according to this
approach, the force acting on a particle at a given moment turns out to
be affected by the entire {\em future} evolution of the particle.

A formal method for calculating the momentary
self force in curved spacetime, which employs the purely retarded Green's
function and is thus inherently causal, was developed long ago by
DeWitt and Brehme\cite{DB} (for the electromagnetic case).
Dewitt and Brehme first carried out
a local analysis of the retarded Green's function near the particle's
worldline, based on the {\em Hadamard expansion} \cite{Hadamard}
(which is, basically, an expansion of the Green's function in powers
of the geodesic distance between the source and evaluation points).
Then, the particle's equation of motion was deduced by imposing local
energy-momentum conservation on a thin world-tube surrounding the
particle's worldline.
Following the same approach (though using a different renormalization
scheme), Mino, Sasaki, and Tanaka \cite{MST} studied the {\em
gravitational} self force by analyzing the local metric perturbations
near a particle. They concluded that the regularized gravitational
self force in vacuum is due solely to a nonlocal tail contribution.
It remained unclear, though, how to practically evaluate the formal
expression derived for this tail (see, however, recent attempts to
tackle this problem \cite{Mino}).

Recently, Quinn and Wald \cite{QW1} developed a different, axiomatic,
approach to the regularization problem of the self force in curved
spacetimes. Their approach relies on a ``comparison axiom'', which allows
the calculation of the self force
by comparing the given problem with a (properly chosen)
analogous problem in flat space (see \cite{QW1} for details).
The implementation of this approach for both the electromagnetic
and the gravitational cases \cite{QW1}
yielded formal results in agreement with those obtained by
DeWitt and Brehme (as corrected by Hobbs \cite{Hobbs}; the main result
by DeWitt and Brehme contained a trivial error)
and by Mino {\em et al.}, with the advantage of involving much simpler
calculations. More recently, the same approach was applied by Quinn
\cite{Quinn} for a scalar particle [the main results of this work,
also quoted in Ref.\ \cite{QW2}, are summarized below;
see Eqs.\ (\ref{eq2-90})--(\ref{eq2-130})].
Again, however, despite the availability of a simple formal framework
for obtaining equations of motion for a test particle in curved
spacetime, the practical implementation of the formalism in actual
calculations (particularly, the evaluation of the nonlocal tail
contribution) remained a challenging task.

So far, the study of the self force effect in concrete situations
have been restricted to very few simple cases.
DeWitt and DeWitt \cite{DD} employed the above mentioned formalism by
DeWitt and Brehme to study the self force correction to the
geodesic equation for an electrically charged particle freely falling
in a static weak gravitational field, in the limit of small velocity.
They concluded that a repulsive force of $\propto q^2r^{-3}$ magnitude
(where $r$ is the Schwarzschild radial coordinate) would be experienced
by such a particle, in addition to the usual attractive inverse-square
force.
Later, Smith and Will \cite{SW} (and, independently, Frolov and
Zel'nikov \cite{FZ1}) were able to derive an exact analytic expression
for the $O(q^2)$ self force acting on an electrically charged
particle held static in the Schwarzschild exterior. They found
a repulsive self force of exact magnitude $Mq^2r^{-3}$
[measured by a momentarily static freely falling observer, and
expressed in relativistic (geometrized) units].
This result was later extended by Frolov and Zel'nikov \cite{FZ2} to
scalar and electrically-charged particles held static outside a
charged, Reissner-Nordstr\"{o}m type, black hole. It was concluded that,
unlike in the electromagnetic case, no self force is experienced by a
static scalar particle. This last result has been reproduced very recently
by Wiseman \cite{Wiseman} in a thorough analysis of the self force
acting on a static scalar particle in Schwarzschild spacetime.

The exact calculation in the static electric charge case was made
possible owing to the existence of an exact analytic solution,
discovered long ago by Copson \cite{Copson} (and later corrected by
Linet \cite{Linet}), for the electrostatic potential of a static charge
in the Schwarzschild geometry. (The analogous closed-form solution for
a scalar particle was constructed by Wiseman in \cite{Wiseman}.)
In more general cases one cannot benefit
from the existence of such exact solutions. The usual approach for
treating this problem, of obtaining solutions to the field equation
in black hole backgrounds, is through the Fourier-multipole
decomposition of the field. In the context of the radiation reaction
problem this approach seems to offer two obvious advantages:
First, it allows, in the usual manner, reduction of the field
equation [originally a partial differential equation (DE) in 1+3
dimensions] to an ordinary DE, thus making it accessible to
simple numerical treatment. Second, each individual mode of the field
turns out to be continuous (and the corresponding self force to be
bounded) even at the particle's location.
Having this in mind, Ori previously proposed \cite{Ori}
that a practical calculation of the self force effect may be carried
out by first evaluating the effect of each Fourier-multipole mode
of a particle's self retarded field on its radiative evolution
(through the local self force experienced by the particle), and then
summing over all modes.

In Ref.\ \cite{Ori} the above sum-over-modes approach has been successfully
applied for the calculation of the adiabatic, orbit-integrated, evolution
rate of the three constants of motion in Kerr spacetime, including
the Carter constant.\footnote
{However, whereas the mode sum for the evolution rate of the energy and
azimuthal angular momentum parameters was shown to converge \cite{Ori},
it is not clear yet whether the corresponding mode sum for the Carter
constant converges or not.}
However, it appears that a naive application of this method for the
calculation of the {\em momentary} self force would not be useful,
in general. The reason is that, although each mode yields a finite
contribution to the self force, the sum over all modes is found, in general,
to diverge. This situation manifests itself already in the most simple
case, that of a static scalar charge in flat space: in this
basic example, the
contribution of each multipole mode to the radial component of the
self force is the same, $-q^2/(2r_0^2)$ (where $r_0$ is the distance
of the particle from the origin of coordinates, with respect to which the
spherical harmonic functions are defined), with an obvious divergence
of the sum over modes.
This, however, does not mean that one has to abandon the whole
sum-over-modes approach; one may still be able to benefit from its
advantages, by introducing a suitable regularization procedure
into the calculation, properly designed to overcome the above kind
of divergence.

In a previous paper \cite{Rapid} we introduced the basic elements of a
method for the calculation of the self force in curved spacetime
through regularization of the mode sum.
The implementation of the proposed calculation method for a specific
trajectory in a given spacetime consists of two stages.
First, one solves (numerically, in most cases) the appropriate ordinary
DE for each Fourier-harmonic mode $\l,m,\omega$ of the retarded field,
and evaluates the (finite) contribution of each of these modes to
the self force. (Alternatively, one may numerically solve the 1+1
partial DE in the time domain, for each multipole mode $l,m$.)
Then, the sum over all modes is made subject to a certain regularization
procedure, which requires the knowledge of several regularization parameters.
These parameters are derived analytically, for any given trajectory,
through local perturbative analysis of the (retarded) Green's function.
In Ref.\ \cite{Rapid} we outlined this regularization method as applied
to a scalar particle moving on a Schwarzschild background, and presented
final results (i.e., the values of all necessary regularization
parameters) for the case of radial motion.
The target of the present paper is three-fold:
(i) providing a systematic presentation of the regularization scheme
(including a discussion of some mathematical subtleties left untreated
in \cite{Rapid});
(ii) providing full details of the calculations involved in deriving
the regularization parameters for radial trajectories; and
(iii) extending the analysis to a wider class of static
spherically-symmetric black hole spacetimes.

This paper (as well as Ref.\ \cite{Rapid}) is concerned with the
{\em analytic} part of the regularization scheme; namely, it sets
the mathematical foundation for the scheme, and demonstrates
the calculation of the regularization parameters involved
in its implementation (in the example of radial motion).
As explained above, full calculation of the self force requires
the supplementary numerical determination of the various modes'
``bare'' contributions to the self force.
This was recently done for various trajectories of a scalar
particle outside a Schwarzschild black hole: Burko first analyzed
the case of a static particle \cite{Burko1} and the one of a particle
in circular motion \cite{Burko2}(see also \cite{Amaldi}).
More recently, Barack and Burko applied the regularization scheme
for studying radial trajectories in Schwarzschild \cite{BB}.
These numerical works confirm the applicability of the regularization
scheme, and provide support for the values of the analytically
deduced regularization parameters. Of course, they also yield
significant physical information.
In the static scalar particle case, Burko recovered the familiar
result, of a zero self force. Calculations of the self force on
a scalar particle in circular and radial trajectories were carried
out for the first time
(see Refs.\ \cite{Burko1,Burko2,BB} for details).

The current paper is organized as follows: In Sec.\ \ref{secII} we give
some preliminary relations involving the self field, the Green's
function, and the self force for a scalar particle. In Sec.\ \ref{secIII}
we decompose the Green's function into its spherical harmonic
components, and discuss the applicability of this expansion.
In Sec.\ \ref{secIV} we decompose the (tail part of the) self force into
its spherical harmonic contributions, discuss the need for
regularization of the mode sum, and present the regularization scheme.
The implementation of this scheme involves local analysis of the Green's
function modes for large multipole numbers, which is carried out
in Sec.\ \ref{secV}. The particular case of radial motion is then
considered in Secs.\ \ref{secVI} and \ref{secVII}, where the regularization
parameters for this case are explicitly calculated.
In Sec.\ \ref{secVIII} we summarize, discuss possible extensions of the
analysis, and briefly survey some related work.

\section{Self force on a scalar charge: preliminaries}\label{secII}

We consider a class of static spherically-symmetric (not necessarily
vacuum) black hole geometries, having a line element of the form\footnote{
Throughout this paper we use relativistic units (with $G=c=1$),
and metric signature ${-}{+}{+}{+}$.}
\begin{equation}\label{eq2-10}
ds^2=-f(r)dt^2+f^{-1}(r)dr^2+r^2(d\theta^2+\sin^2\theta\, d\varphi^2),
\end{equation}
where $t$, $r$, $\theta$, and $\varphi$ are the Schwarzschild coordinates,
and $f$ is a function of $r$ only, positive outside the event horizon.
Important members of this class include the Schwarzschild solution,
with $f=1-2M/r$; the Reissner-Nordstr\"{o}m solution, with
$f=1-2M/r+Q/r^2$; and the Schwarzschild-de Sitter solution, with
$f=1-2M/r+\Lambda r^2/3$. Here, $M$ stands for the black hole's mass, $Q$
represents its net electric charge, and $\Lambda$ is the cosmological
constant.

We next consider a point-like particle of scalar charge $q$, with
$|q|\ll M$, moving in a
spacetime of the above type. Let $x^{\mu}=x_p^{\mu}(\tau)$ represent the
particle's worldline (not necessarily a geodesic), with $\tau$ being its
proper time. The scalar particle exhibits a Klein-Gordon field $\Phi$,
satisfying
\begin{equation}\label{eq2-20}
\Box\Phi\equiv \Phi_{;\alpha}{^{;\alpha}}=
\frac{1}{\sqrt{-g}}\left(\sqrt{-g}\,g^{\alpha\beta}
\Phi_{,\alpha}\right)_{,\beta}=-4\pi\rho(x^{\mu}),
\end{equation}
where $\Box$ represents the covariant D'Alembertian operator, $g$ is the
metric determinant, and $\rho(x^{\mu})$ is the scalar charge density,
given by
\begin{equation}\label{eq2-30}
\rho(x^{\mu})= q \int_{-\infty}^{\infty}\frac{1}{\sqrt{-g}}\,
\delta^4(x^{\mu}-x_p^{\mu}(\tau))d\tau.
\end{equation}
The solution for the scalar field can be written as
\begin{equation}\label{eq2-40}
\Phi(x^{\mu})=\int G(x^{\mu};{x'}^{\mu})\rho({x'}^{\mu})
{\sqrt{-g}}\, d^4 x',
\end{equation}
where $G(x^{\mu};{x'}^{\mu})$ is the retarded Green's function, satisfying
\begin{equation}\label{eq2-50}
\Box G(x^{\mu};{x'}^{\mu})=\frac{-4\pi}{\sqrt{-g}}\,
\delta^4(x^{\mu}-{x'}^{\mu}),
\end{equation}
and subject to the causality condition, $G=0$ whenever
$x^{\mu}$ lies outside the future light cone of ${x'}^{\mu}$.
Combining Eqs. (\ref{eq2-30}) and (\ref{eq2-40}) we obtain for
the scalar field
\begin{equation}\label{eq2-60}
\phi(x^{\mu})= q\int_{-\infty}^{\infty}
G\left[x^{\mu};x_p^{\mu}(\tau)\right] d\tau.
\end{equation}

The ``scalar force'' experienced by the particle due to its own field
shall be taken, following \cite{QW2}, to be
\begin{equation}\label{eq2-70}
F_{\alpha}=q\Phi_{;\alpha}=q\Phi_{,\alpha},
\end{equation}
evaluated at the particle's location.
We comment that the so-defined force is not perpendicular to the
four-velocity of the particle, $u^{\alpha}\equiv dx_p^{\alpha}(\tau)/d\tau$,
resulting in that the mass parameter of the particle is not conserved along
the worldline. Indeed, the force on a scalar particle can be calculated
otherwise (as in \cite{Gal'tsov}, e.g.), such as to make the mass parameter
conserved:
$F_{\alpha}^{\bot}=q(\Phi_{,\alpha}+u_{\alpha}u^{\beta}\Phi_{,\beta})$.
Although we shall adopt here the simpler definition, Eq.\ (\ref{eq2-70}),
the results of our analysis could then easily be applied for the
force $F_{\alpha}^{\bot}$ as well. (Given all vectorial components of
$F_{\alpha}$, one can easily construct both the force component
perpendicular to the worldline, and the component tangent to it.)
With Eq.\ (\ref{eq2-60}) we now have for the self force acting on the
particle at a point $x_0^{\mu}\equiv x_p^{\mu}(\tau=0)$ along its worldline,
\begin{equation}\label{eq2-80}
F_{\alpha}= q^2\int_{-\infty}^{\infty}
G_{,\alpha}\left[x^{\mu};x_p^{\mu}(\tau)\right] d\tau,
\end{equation}
where the gradient (taken with respect to $x^{\mu}$)
is to be evaluated at $x^{\mu}=x_0^{\mu}$.

The ``bare'' self force given in Eq.\ (\ref{eq2-80})
needs to be regularized to avoid divergences associated with the behavior
of the scalar field at the very location of the particle.
For that goal, Quinn \cite{Quinn,QW2} applied the ``comparison axiom''
approach by Quinn and Wald \cite{QW1} for the scalar particle case.
The total self force acting on the scalar particle was found to be
composed of three parts:
\begin{equation}\label{eq2-90}
F^{\rm (total)}_{\alpha}=F^{\rm (ALD)}_{\alpha}+F^{\rm (Ricci)}_{\alpha}
+F^{\rm (tail)}_{\alpha}.
\end{equation}
The first term here is a local ALD-like term,
reading
\begin{equation}\label{eq2-100}
F^{\rm (ALD)}_{\alpha}=\frac{1}{3}q^2(\dot{a}_{\alpha}-a^2 u_{\alpha}),
\end{equation}
where $a^{\alpha}$ is the four-acceleration of the particle,
$a^2\equiv a_{\beta}a^{\beta}$,
and an overdot represents covariant differentiation with respect to
the particle's proper time $\tau$.
The second term in Eq.\ (\ref{eq2-90}) is related to the local Ricci
curvature at the particle location. It is given by
\begin{equation}\label{eq2-110}
F^{\rm (Ricci)}_{\alpha}=\frac{1}{6}q^2(R_{\alpha\beta}u^{\beta}+
u_{\alpha}R_{\beta\gamma}u^{\beta}u^{\gamma}-Ru_{\alpha}/2),
\end{equation}
where $R_{\alpha\beta}$ is the Ricci tensor and $R$ is the curvature
scalar.
The third term in the expression for the total self force represents
the non-local ``tail'' contribution. It may be expressed as
\begin{equation}\label{eq2-120}
F^{\rm (tail)}_{\alpha}\equiv\lim_{\epsilon\to 0^+}F^{(\epsilon)}_{\alpha},
\end{equation}
where
\begin{equation}\label{eq2-130}
F^{(\epsilon)}_{\alpha}\equiv q^2\int_{-\infty}^{-\epsilon}
G_{,\alpha}\left[x^{\mu}_0;x_p^{\mu}(\tau)\right] d\tau.
\end{equation}
As we mentioned above, the occurrence of a tail term --- a prominent
feature of the self force in curved spacetimes ---
is due to the Green's function having its support also {\em inside}
the source's future light cone. From the physical point of view, this is
associated with the fact that waves are scattered off spacetime curvature
while propagating on a curved background.

The task of implementing the formal expression, Eq.\ (\ref{eq2-90}),
in practical calculations of the self force is a challenging one.
The difficulty stems, of course, from the need to evaluate the tail part,
which requires knowledge of the Green's function everywhere along
the particle's past worldline.
Below we therefore focus on the tail term contribution to the self force,
presenting a practical method for its calculation.

\section{Multipole decomposition of the Green's function}\label{secIII}

The regularization scheme to be introduced below is based on evaluating
the contribution to the (tail part of the) self force due to each multipole
mode of the (retarded) Green's function. To that end we first consider the
multipole decomposition of the Green's function.

To begin, one may be tempted to decompose $G$ into its multipole modes
$G^l$ in the usual manner, as $G=\sum_{l=0}^{\infty} G^{l}$
(where $G^{l}$ represents the quantity resulting from summing
over azimuthal numbers $m$). Although this may look
as standard procedure, caution is necessary here: in general, such a
decomposition turns out ill-defined, as the sum over $l$ is found to diverge.
This can be illustrated already in flat space. In this case, the modes
$G^l$ admit a closed-form expression, which, for evaluation point $x^{\mu}$
lying inside the future light cone of the source point ${x'}^{\mu}$,
is given by
\begin{equation}\label{eq3-4}
G^l_{\rm Flat}(x^{\mu},{x'}^{\mu})=
\frac{(2l+1)P_l(\cos\chi)P_l[1-\sigma/(rr')]}{2rr'}.
\end{equation}
Here, $P_l$ is the Legendre polynomial,
$\sigma\equiv \frac{1}{2}\left[(t-t')^2-(r-r')^2\right]$, and
\begin{equation}\label{eq3-6}
\cos\chi\equiv\cos\theta \cos\theta' + \sin\theta \sin\theta'
\cos(\varphi-\varphi').
\end{equation}
[Eq.\ (\ref{eq3-4}) can be verified by direct substitution, using
Eqs.\ (\ref{eq3-110}), (\ref{eq3-120}), and (\ref{eq3-150}), to be
given below.]
Consider, for example, the case $\chi=0$, with $P_l(\cos\chi)\equiv 1$
for all $l$, corresponding to both the source and evaluation points
lying in the same radial direction.
At large $l$ values, the Legendre polynomial $P_l(\sigma)$ admits the
asymptotic form
$\propto l^{-1/2}\times$ oscillations with respect to $l$
[the exact asymptotic form is given in App.\ \ref{AppA} below; see
Eq.\ (\ref{eqA-150})].
Thus, at large $l$ one finds $G^l_{\rm Flat}\propto l^{1/2}\times$
oscillations, implying that the infinite sum over all modes
fails to converge.\footnote{In the more general case, with $\chi\ne 0$,
there is also a $\propto l^{-1/2}\times {\rm oscillations}$ factor coming
at large $l$ from $P_l(\cos\chi)$, yielding the asymptotic form
$G^l_{\rm Flat}\propto {\rm const\ }\times$ oscillations. Hence, clearly,
the sum over modes $G^l$ diverges in the general case as well.}
Mathematically speaking, this failure of the naive multipole decomposition
may be associated with the fact that the Green's function
of our problem exhibits a strong irregularity along the intersection
of the future light cone of the source with the sphere of constant $r$
and $t$. In Appendix \ref{AppA} we discuss this irregularity in more
detail, referring to the analysis by DeWitt and Brehme \cite{DB}.
We then suggest a way to overcome the difficulty caused by the presence
of the irregularity, and explain how a well defined mode decomposition
can still be accomplished. Although the detailed discussion of this issue
is left to the Appendix, we outline here the basic argument, and
present some definitions and notation needed in the sequel.

In Appendix \ref{AppA} we construct a ``modified'' Green's function
$G_{\rm mod}=G-\delta G$, where the function $\delta G$ is chosen such
that $G_{\rm mod}$ has the following properties: (i) it is a continuous
function of $\theta$ and $\varphi$ across the sphere of constant $r$
and $t$; and (ii) it yields the same self force, through Eq.\ (\ref{eq2-130}),
as the original Green's function $G$ (this is guaranteed by taking the
function $\delta G$ to have no support inside the future light cone of
the source). It is then argued that the modified Green's function
admits an (absolutely) convergent multipole expansion,
$G_{\rm mod}=\sum_l (G^l-\delta G^l)$.
Next, we define the new operation $\widetilde\lim_{l\to\infty}$
(``tilde-limit'') of a series of numbers $A_l$, as the standard limit
$\lim_{l\to\infty}$ (when existing and finite) of the series
$A_l-B_l^{(1)}-B_l^{(2)}\cdots-B_l^{(k)}$,
where the $B_l^{(j)}$'s are any finite number of terms having the form
$B_l^{(j)}=a_j l^{b_j}\cos(\alpha_j l+\beta_j)$,
with $a_j$, $b_j$, $\alpha_j$, and $\beta_j$ being
some $l$-independent real numbers, and with none of the numbers
$\alpha_j$ vanishing. Namely, if there exist $k$ quantities $B_l^{(j)}$
of the above form, such that subtracting them from the original series
$A_l$ would yield a well-defined finite limit as $l\to\infty$, then
we define
\begin{equation}\label{eq3-7}
\widetilde\lim_{l\to\infty}A_l\equiv
\lim_{l\to\infty}\left[A_l-\sum_{j=1}^k B_l^{(j)}\right].
\end{equation}
We also define the ``tilde-sum'' of a series $A_l$ by
\begin{equation}\label{eq3-8}
\sum_{l=0}^{\widetilde\infty} A_l\equiv \widetilde\lim_{\bar{l}\to\infty}
\sum_{l=0}^{\bar{l}}A_l,
\end{equation}
where $\sum_l^{\bar l}$ is the standard summation operation.
It can be easily verified (see App.\ \ref{AppA}) that when the
``tilde-limit'' (or the ``tilde-sum'') of a series exists,
then it is {\em unique}. In particular, if a standard infinite sum
$\sum_{l}^{\infty}A_l$ converges, then one may replace it with a
``tilde-sum'' operation. Thus, we may replace the
convergent standard sum $G_{\rm mod}=\sum_{l=0}^{\infty}(G^l-\delta G^l)$
with a tilde-summation. In Appendix \ref{AppA} we show that
$\sum_{l=0}^{\widetilde\infty}(\delta G^l)$=0.
Consequently, we conclude
\begin{equation}\label{eq3-10}
G_{\rm mod}=\sum_{l=0}^{\widetilde\infty}G^l.
\end{equation}
We emphasize once more that the modified Green's function $G_{\rm mod}$
can serve {\em instead} of the original function $G$ for the calculation
of self force, as both functions yield the same force.
Eq.\ (\ref{eq3-10}) thus implies that the calculation of the self force
can be carried out through analysis of the original Green's function's
modes $G^l$, by applying the {\em tilde}-summation instead of the
(ill-defined) standard summation. (A more thorough discussion of the
arguments leading to Eq.\ (\ref{eq3-10}) will be given in Appendix
\ref{AppA}.)

Let us now turn to study the form of the multipole modes $G^l$
in greater detail. These modes can be written more explicitly as
\begin{equation}\label{eq3-20}
G^l(x^{\mu},{x'}^{\mu})= \sum_{m=-l}^{l}
Y_{lm}(\theta,\varphi)\, \hat{g}^{lm}(t,r;{x'}^{\mu}),
\end{equation}
where $Y^{lm}(\theta,\varphi)$ are the standard spherical harmonic
functions on the sphere of constant $r$ and $t$.
Substituting Eq.\ (\ref{eq3-20}) and the relation
$
\delta(\theta-\theta')\delta(\varphi-\varphi')/\sin\theta=
\sum_{l,m}Y_{lm}(\theta,\varphi)Y_{lm}^*(\theta',\varphi'),
$
(where an asterisk denotes complex conjugation) in Eq.\ (\ref{eq2-50}),
we obtain from the orthogonality of the spherical harmonics,
\begin{equation}\label{eq3-40}
r^2f^{-1}(r)\hat{g}_{,tt}^{lm}-\left[r^2 f(r)\hat{g}_{,r}^{lm}
\right]_{,r}  +l(l+1)\hat{g}^{lm}=
4\pi\delta(t-t')\delta(r-r') Y_{lm}^*(\theta',\varphi').
\end{equation}
In terms of the $m$-independent variable $\tilde{g}^{l}(t,r;t',r')$,
defined through
\begin{equation}\label{eq3-50}
\hat g^{lm}=2\pi\tilde{g}^{l}Y_{lm}^*(\theta',\varphi')/(r r'),
\end{equation}
Eq.\ (\ref{eq3-40}) becomes
\begin{equation}\label{eq3-60}
\tilde{g}_{,tt}^{l}- \tilde{g}_{,r_*r_*}^{l} + 4V^{l}(r)
\tilde{g}^{l}= 2f(r)\delta(t-t')\delta(r-r'),
\end{equation}
where the radial coordinate $r_*(r)$ admits $dr_*/dr=f^{-1}(r)$,
and the effective potential $V^{l}(r)$ is given by
\begin{equation}\label{eq3-70}
V^{l}(r)=\frac{1}{4}f(r)
\left(\frac{l(l+1)}{r^2}+\frac{f'(r)}{r}\right)
\end{equation}
(with a prime denoting $d/dr$).

To account for the causality condition, it is convenient to introduce
the (Eddington-Finkelstein-like) null coordinates
\begin{equation}\label{eq3-80}
v\equiv t+r_*\quad\text{and}\quad u\equiv t-r_*.
\end{equation}
The relation
$\delta(t-t')\delta(r-r')=
2f^{-1}(r')\delta(v-v')\delta(u-u')$
can then be used to write Eq.\ (\ref{eq3-60}) in the simple form
\begin{equation}\label{eq3-90}
\tilde{g}_{,vu}^{l}+V^{l}(r)\tilde g^{l}=\delta(v-v')\delta(u-u').
\end{equation}
We now impose causality by writing
\begin{equation}\label{eq3-100}
\tilde g^{l}=g^{l}(v,u;v',u') \Theta(v-v')\Theta(u-u'),
\end{equation}
where $\Theta$ is the standard step function.
The ``reduced Green's function'' $g^{l}(v,u;v',u')$ obeys
the homogeneous equation
\begin{equation}\label{eq3-110}
g_{,vu}^{l}+V^{l}(r)g^{l}=0
\end{equation}
for all $u>u'$ and $v>v'$.
Substituting Eq.\ (\ref{eq3-100}) into Eq.\ (\ref{eq3-90}) and examining
the behavior along the null rays $v=v'$ and $u=u'$, one
finds that $g^{l}$ must admit
\begin{equation}\label{eq3-120}
g^{l}(v=v')=g^{l}(u=u')=1.
\end{equation}
For any fixed source point $v',u'$, the homogeneous equation
(\ref{eq3-110}), supplemented by the initial conditions (\ref{eq3-120}),
constitutes a characteristic initial-value problem for
the function $g^{l}$ anywhere at $u>u'$ and $v>v'$.

Finally, to express $G^{l}$ in terms of the reduced Green's function
$g^{l}$, we substitute Eq.\ (\ref{eq3-50}) [with Eq.\ (\ref{eq3-100})]
into Eq.\ (\ref{eq3-20}). In the resulting expression we can explicitly
sum over $m$ by making use of the relation \cite{Jackson}
\begin{equation}\label{eq3-130}
\sum_{m=-l}^{l}Y_{lm}(\theta,\varphi)Y_{l m}^*(\theta',\varphi')
=(4\pi)^{-1}(2l+1)P_{l}(\cos\chi),
\end{equation}
where $\cos\chi$ is the quantity given in Eq.\ (\ref{eq3-6}).
We then find for the $l$-mode of the Green's function,
\begin{equation}\label{eq3-150}
G^{l}= L P_{l}(\cos\chi)
\frac{g^{l}(v,u;v',u')}{r r'}\, \Theta(v-v')\,\Theta(u-u'),
\end{equation}
where we have set
\begin{equation}\label{eq3-160}
L\equiv l+1/2.
\end{equation}

\section{Mode sum regularization scheme}\label{secIV}

\subsection{The need for a mode sum regularization}

Following the discussion of the preceding section, we now replace $G$
in Eq.\ (\ref{eq2-130}) by $G_{\rm mod}$, and then substitute
for $G_{\rm mod}$ from Eq.\ (\ref{eq3-10}). We find\footnote{
It is assumed here that both the differentiation and the integration
involved in constructing $F_{\alpha}^{(\epsilon)}$ out of $G$ can
be performed term-by-term with respect to the tilde-summation.
This assumption should be verified by more closely inspecting the
convergence properties of the tilde-sum over $G^l$ in Eq.\
(\ref{eq3-10}), which, however, would be beyond the scope of the
current paper.}
\begin{eqnarray}\label{eq4-10}
F_{\alpha}^{(\epsilon)}= \sum_{l=0}^{\widetilde\infty}
F^{l(\epsilon)}_{\alpha},
\end{eqnarray}
where $F^{l(\epsilon)}_{\alpha}$ represents the contribution to
$F_{\alpha}^{(\epsilon)}$ associated with the $l$-mode Green's
function:
\begin{equation}\label{eq4-20}
F^{l(\epsilon)}_{\alpha}=q^2\int_{-\infty}^{-\epsilon}
G^{l}_{,\alpha}\left[x_0^{\mu};x_p^{\mu}(\tau)\right] d\tau.
\end{equation}

For practical reasons which become clear below, let us now write
\begin{equation}\label{eq4-30}
F^{l(\epsilon)}_{\alpha}=
F^{l}_{\alpha}-\delta F^{l(\epsilon)}_{\alpha},
\end{equation}
in which\footnote{
Strictly speaking, the two quantities $F^{l}_{\alpha}$ and
$\delta F^{l(\epsilon)}_{\alpha}$ are not well defined without
specifying the direction through which the gradient of $G^{l}$ is
calculated. This issue is discussed in length later in this section.}
\begin{equation}\label{eq4-40}
F^{l}_{\alpha}=q^2\int_{-\infty}^{\infty}
G^{l}_{,\alpha}\left[x_0^{\mu};x_p^{\mu}(\tau)\right] d\tau
\quad \text{and}\quad
\delta F^{l(\epsilon)}_{\alpha}=q^2\int_{-\epsilon}^{\infty}
G^{l}_{,\alpha}\left[x_0^{\mu};x_p^{\mu}(\tau)\right] d\tau.
\end{equation}
Here, $F^{l}_{\alpha}$ is the $l$-mode of
$F_{\alpha}=q\Phi_{,\alpha}$---the quantity given in Eq.\ (\ref{eq2-80}),
which is sourced by the entire worldline. This quantity can be obtained
from the $l$-mode of the self field, which, in turn, can be calculated
essentially with no difficulty (using numerical methods, in most cases
\cite{Burko1,Burko2,BB}). The other quantity appearing in Eq.\
(\ref{eq4-30}), $\delta F^{l(\epsilon)}_{\alpha}$, is local in nature,
and thus may be treated, in principle, by means of local analytic
methods (as we, indeed, demonstrate in this paper).

In terms of $F^{l}_{\alpha}$ and $\delta F^{l(\epsilon)}_{\alpha}$,
the tail part of the self force is calculated through
\begin{equation}\label{eq4-50}
F^{\rm (tail)}_{\alpha}=
\lim_{\epsilon\to 0^+}\sum_{l=0}^{\widetilde\infty}\left(F^{l}_{\alpha}
-\delta F^{l(\epsilon)}_{\alpha} \right).
\end{equation}
To carry out this calculation, one may tempt to first calculate
the sum over $F^{l}_{\alpha}$ (which is $\epsilon$-independent),
and then evaluate the local contribution
$\lim_{\epsilon\to 0^+}\widetilde\sum_{l} \delta F^{l(\epsilon)}_{\alpha}$.
However, here one comes across a problem:
Although each of the modes $F^{l}_{\alpha}$ yields a finite contribution
at the particle's location, in general {\em the sum over all modes
$F^{l}_{\alpha}$ diverges}. As we mentioned in the Introduction, this can
be demonstrated even in the simple case of a static scalar charge in flat
space. To overcome this type of divergence, the introduction of a certain
regularization procedure for the mode sum is required.
Such a procedure is described (and later implemented) in what follows.

\subsection{The regularization scheme}

To regularize the modes sum, one seeks a (simple as possible)
$\epsilon$-independent function $h^{l}_{\alpha}$, such that the
series $\sum_{l}(F^{l}_{\alpha} - h^{l}_{\alpha})$ would converge.
Once such a function is found, Eq.\ (\ref{eq4-50}) can be written as
\begin{equation}\label{eq4-60}
F^{\rm (tail)}_{\alpha}=\sum_{l=0}^{\infty}\left(F^{l}_{\alpha}
-h^{l}_{\alpha} \right) - D_{\alpha},
\end{equation}
where
\begin{equation}\label{eq4-70}
D_{\alpha} \equiv \lim_{\epsilon\to 0^+}\sum_{l=0}^{\widetilde\infty}
\left(\delta F^{l(\epsilon)}_{\alpha} - h^{l}_{\alpha}\right).
\end{equation}

In principle, a regularization function $h^{l}_{\alpha}$ can be
constructed by investigating the asymptotic behavior of $F^{l}_{\alpha}$
as $l \to \infty$.
It is also possible, however, to derive $h^{l}_{\alpha}$ from the
large-$l$ asymptotic behavior of $\delta F^{l(\epsilon)}_{\alpha}$:
The latter and $F^{l}_{\alpha}$ must have the same
singular behavior at the tilde-limit $l\to\infty$ (for fixed $\epsilon$),
as their difference yields a convergent tilde-sum over $l$.
Obviously, in order to determine $h^{l}_{\alpha}$ (and $D_{\alpha}$)
from $\delta F^{l(\epsilon)}_{\alpha}$,
one merely needs the asymptotic behavior of
$\delta F^{l(\epsilon)}_{\alpha}$ in the immediate neighborhood of
$\epsilon=0$. This allows one to derive $h^{l}_{\alpha}$ (and
$D_{\alpha}$) using local analytic methods, as shall be demonstrated in
the next section.

First, however, it would be necessary to comment here about
a certain indefiniteness involved in the above definitions of the
quantities $F^{l}_{\alpha}$ and $\delta F^{l(\epsilon)}_{\alpha}$.

\subsection*{Discontinuity of $F^{l}_{\alpha}$ and
            $\delta F^{l(\epsilon)}_{\alpha}$}

\begin{figure}[htb]
\input{epsf}
\centerline{\epsfysize 4.5cm \epsfbox{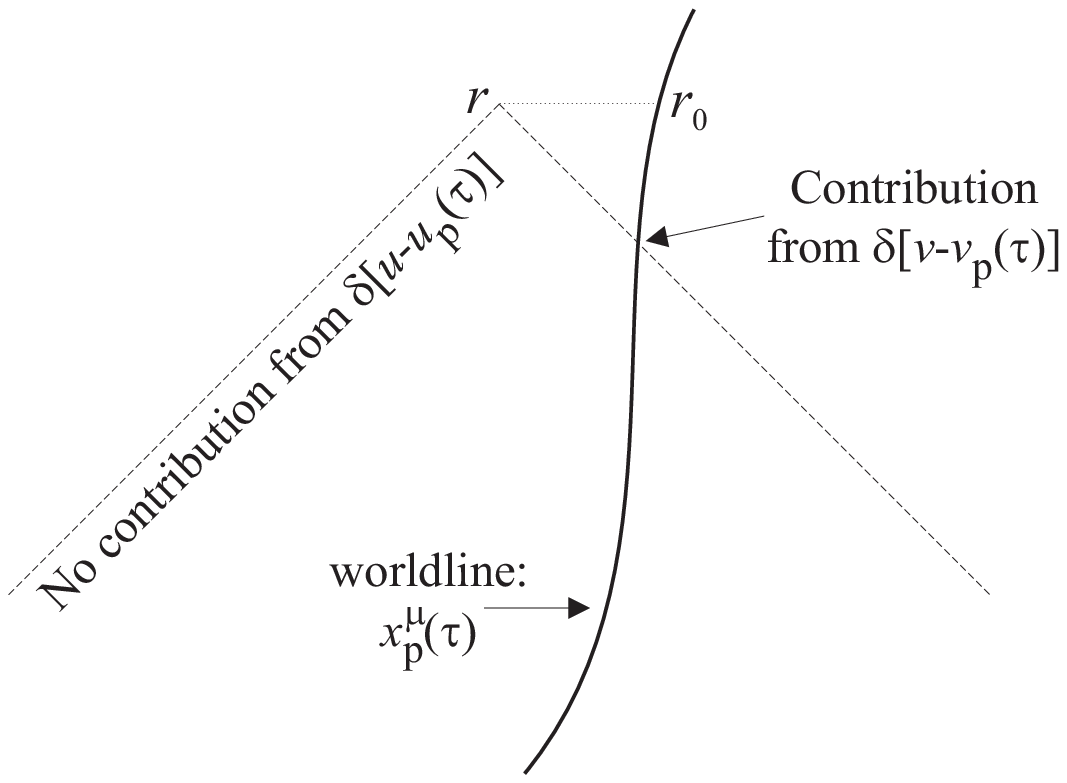}
            \epsfysize 4.5cm \epsfbox{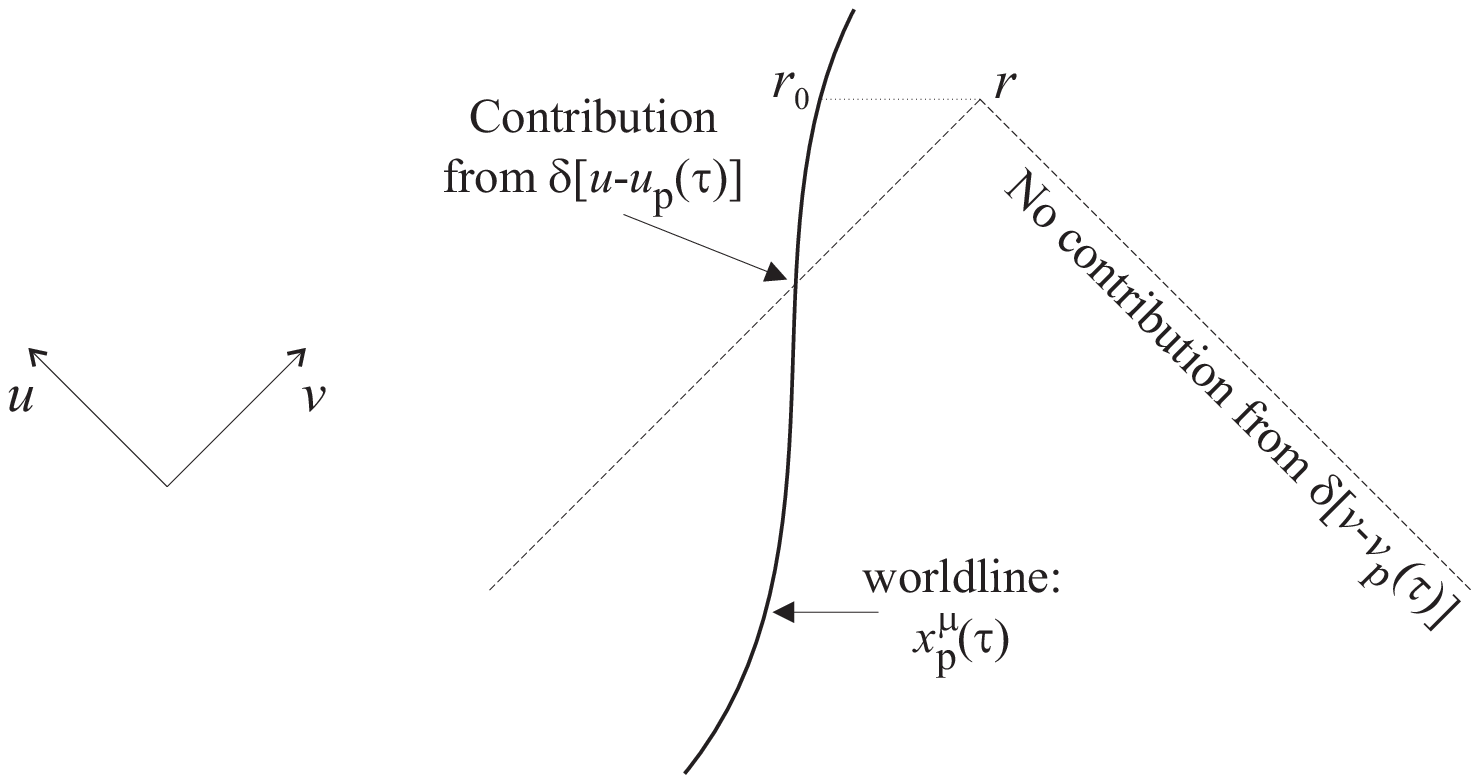}}
\caption{\protect\footnotesize
Discontinuity of  $F^{l}_r$ and $\delta F^{l(\epsilon)}_r$.
These quantities are calculated from Eq.\ (\ref{eq4-40}) by
integrating over $\tau$ the Green's function's $r$-derivative at
the particle's location, $r=r_0$.
If this $r$-derivative is calculated from the limit $r\to r_0^+$
(right figure), then the term in Eq.\ (\ref{eq4-80}) involving
$\delta(u-u_p)$ will contribute to the $\tau$ integration, whereas
the $\delta(v-v_p)$ term will have no contribution.
The situation is reversed if the $r$-derivative is calculated from the
limit $r\to r_0^-$ (left figure): then, contribution will come only
from the $\delta(v-v_p)$ term. Since these two contributions are
different, one finds that each of the quantities $F^{l}_r$ and
$\delta F^{l(\epsilon)}_r$ has two different one-sided values
(both of which are finite in magnitude).}
\label{fig1}
\end{figure}

Whereas the quantity $F^{l(\epsilon)}_{\alpha}$ of Eq.\ (\ref{eq4-20})
is well defined, the values of the two quantities
$F^{l}_{\alpha}$ and $\delta F^{l(\epsilon)}_{\alpha}$
depend on how exactly one evaluates the gradient $G^{l}_{,\alpha}$ at
the particle's location. To make this point clear, consider first the
$r$-components $F^{l}_r$ and $\delta F^{l(\epsilon)}_r$.
These are calculated according to Eq.\ (\ref{eq4-40}) from the
$r$-derivative of $G^{l}$, reading
\begin{eqnarray}\label{eq4-80}
G^{l}_{,r}= \frac{L P_{l}(\cos\chi)}{r r'}
\left\{\left[g^{l}_{,r}-g^{l}/r\right]\Theta(v-v')\Theta(u-u')+
f^{-1}g^{l}\left[\delta(v-v')\Theta(u-u')-\Theta(v-v')
\delta(u-u')\right]\right\}.
\end{eqnarray}
To calculate $F^{l}_r$ and $\delta F^{l(\epsilon)}_r$
one needs to evaluate this derivative at the self
force's evaluation point, $x^{\mu}=x_0^{\mu}$, with a source point
${x'}^{\mu}=x_p^{\mu}(\tau)$. Now, if the derivative at $x_0^{\mu}$ is
calculated from $r_0^+$ [namely, by taking the limit $r\to r_0^+$ of
$\frac{G^{l}(r)-G^{l}(r_0)}{r-r_0}$], then the $\delta(u-u')$ term
in Eq.\ (\ref{eq4-80}) will have a nonvanishing contribution to
$F^{l}_r$ and to $\delta F^{l(\epsilon)}_r$
[through the integrals in Eq.\ (\ref{eq4-40})], whereas the $\delta(v-v')$
term will have no contribution --- see figure \ref{fig1}.
On the other hand, if the derivative is taken from $r^-$, it will be
the $\delta(v-v')$ term to contribute, and the $\delta(u-u')$
term to have no contribution. One can easily verify (as we explicitly do
in the following section) that these two different  $\delta$ terms
yield different contributions to the integrals in Eq.\ (\ref{eq4-40}).
Thus, although each of the quantities $F^{l}_r$ and
$\delta F^{l(\epsilon)}_r$ has well defined values when calculated from
either the limit $r\to r_0^-$ or the limit $r\to r_0^+$, {\em these
two one-sided values do not coincide}.
[Note that the quantity $F^{l(\epsilon)}_r$ defined in Eq.\
(\ref{eq4-20}) does not exhibit this kind of discontinuity, as for any
finite $\epsilon$ neither of the two $\delta$ terms contribute to this
quantity.]

One can similarly show that the $t$-components $F^{l}_t$ and
$\delta F^{l(\epsilon)}_t$ also exhibit this kind of discontinuity
through the particle's location (see the explicit calculation carried
out in section \ref{secV} below).
On the other hand, the $\theta$ and $\varphi$ components are obviously
continuous through the particle's location, as $G^l$
depends on the angular coordinates only through the regular function
$P_{l}(\cos\chi)$.

For the sake of definiteness, we shall denote by $F^{l+}_{r}$
and $\delta F^{l(\epsilon)+}_{r}$ the one-sided values arising from the
$r\to r_0^+$ limit, and by $F^{l-}_r$
and $\delta F^{l(\epsilon)-}_r$ the ones arising from the
$r\to r_0^-$ limit.
In addition, the symbols $F^{l\pm}_t$ and $\delta F^{l(\epsilon)\pm}_t$
will stand for the values derived from the limit $t\to t_0^\mp$
if $dr/dt>0$ at the force's evaluation point, or from the limit
$t\to t_0^\pm$ if $dr/dt<0$ there.
(In case $dr/dr=0$ at the force's evaluation point, the two one-sided
values of the $t$-component turn out to coincide, as we obtain below.)
With this notation we find, for $\alpha=r$ or $t$,
\begin{equation}\label{eq4-90}
\delta F^{l(\epsilon)\pm}_{\alpha}=q^2\int_{-\epsilon}^{0^+}
G^{l\pm}_{,\alpha}\left[x_0^{\mu};x_p^{\mu}(\tau)\right] d\tau
\end{equation}
(and similarly for $F^{l\pm}_{\alpha}$), where
\begin{equation}\label{eq4-100}
G^{l\pm}_{,r}(x^{\mu};{x'}^{\mu})\equiv \frac{L P_{l}(\cos\chi)}{r r'}
\left[g^{l}_{,r}-g^{l}/r \mp f^{-1}g^{l}
\delta(w_{\pm}-w'_{\pm}) \right],
\end{equation}
and
\begin{equation}\label{eq4-110}
G^{l\pm}_{,t}(x^{\mu};{x'}^{\mu})\equiv \frac{L P_{l}(\cos\chi)}
{r r'} \left[g^{l}_{,t}+ g^{l} \delta(w_{\pm}-w'_{\pm}) \right].
\end{equation}
Here, we have introduced the notation
\begin{equation}\label{eq4-120}
w_+\equiv u \quad\text{and}\quad w_-\equiv v,
\end{equation}
and, likewise, $w'_+\equiv u'$ and $w'_-\equiv v'$.


\section{Local analysis of $G^{\lowercase{l}}$ for large
        $\lowercase{l}$}\label{secV}

The execution of the regularization procedure introduced above
involves the construction of the quantities $h^{l}_{\alpha}$ and
$D^{l}_{\alpha}$.
As we pointed out earlier, this can be done by analyzing
$\delta F^{l(\epsilon)}_{\alpha}$, or, more accurately,
the quantities $\delta F^{l(\epsilon)\pm}_{\alpha}$ given
in Eq.\ (\ref{eq4-90}). For that goal, we must have sufficient
information about the Green's function's $l$-mode $G^{l}$,
for large values of $l$, at the immediate vicinity of the self
force evaluation point. In this section we use local analysis
to obtain analytic approximation for $G^{l}$, up to the accuracy
needed for the derivation of $h^{l}_{\alpha}$ and $D^{l}_{\alpha}$.

\subsection{perturbation analysis}

In Sec.\ \ref{secIII} above we have reduced the problem of calculating
$G^{l}$ to that of solving a $(1+1)$-dimensional homogeneous partial
DE for the function $g^{l}$, Eq.\ (\ref{eq3-110}), with the
characteristic initial
data specified in Eq.\ (\ref{eq3-120}). Given the function $g^{l}$,
the ``four-dimensional'' Green's function $l$-mode, $G^{l}$, is
then constructed from Eq.\ (\ref{eq3-150}).

To explore the behavior of the function $g^{l}$ for small spacetime
intervals and large $l$, we apply the following perturbation analysis.
Let us separate the effective potential given in Eq.\ (\ref{eq3-70})
into two pieces, in the form
\begin{equation}\label{eq5-10}
V^{l}(r)=L^2 V_0(r)+V_1(r),
\end{equation}
where
\begin{equation}\label{eq5-20}
V_0(r)=\frac{f(r)}{4r^2}, \quad\text{and}\quad
V_1(r)=\frac{f(r)}{16r^2}\left[4rf'(r)-1\right].
\end{equation}
Let us next expand $V^l(r)$ in a Taylor series in the small deviation
$r-r_0$ about the particle's location $r=r_0$. It is convenient to take
the small expansion parameter to be $r_*-r_{*0}$, yielding
\begin{equation}\label{eq5-30}
V^{l}(r)=V^{l}(r_0)+ \bar{V^{l}}(r_0)(r_*-r_{*0})+
\frac{1}{2}\bar{\bar{V^{l}}}(r_0)(r_*-r_{*0})^2+\cdots,
\end{equation}
where an overbar denotes $d/dr^*$. Let us also define
\begin{equation}\label{eq5-40}
\Delta_{r}\equiv 2V_{00}^{1/2}L(r_*-r_{*0}),
\end{equation}
where $V_{00}\equiv V_0 (r_0)$.
We shall refer to a variable of this kind, having the form
$L\times\text{(small spacetime deviation)}$, as a ``neutral'' variable.
Such ``neutral'' variables shall play an important role in our analysis,
allowing one to properly take into account the delicate interplay between
large $l$ and small spacetime deviations.
Expressing $r_*-r_{*0}$ in terms of $\Delta_{r}/L$, substituting in the
above Taylor expansion, and collecting terms of the same powers in $L$
(with fixed $\Delta_{r}$), Eq.\ (\ref{eq5-30}) takes the form
\begin{equation}\label{eq5-50}
V^l (r)=V_{00}\left[L^2+
L\left(f_1 \Delta_{r}\right)+ \left(f_2+f_3 \Delta_{r}^2\right)
\right]+O(1/L),
\end{equation}
where $f_1$, $f_2$, and $f_3$ are coefficients given by
\begin{eqnarray}\label{eq5-60}
f_1&\equiv& \frac{1}{2}V_{00}^{-3/2}\bar{V_0}=f^{-1/2}(r f'-2f),
\\
f_2&\equiv&  V_{00}^{-1}V_1 = r f'-1/4,
\\
f_3&\equiv& \frac{1}{8}V_{00}^{-2}\bar{\bar{V_0}}=
\frac{r^2}{2}\left[(f')^2/f+f''\right]+3(f-rf').
\end{eqnarray}
(Here, all quantities are evaluated at $r=r_0$.)

Defining now the dimensionless ``neutral'' coordinates
\begin{equation}\label{eq5-70}
y= V_{00}^{1/2}L(v-v')\quad\text{and}\quad
x= V_{00}^{1/2}L(u-u'),
\end{equation}
Eq.\ (\ref{eq3-110}) becomes
\begin{equation}\label{eq5-80}
g_{,yx}^{l}+\left[1+\frac{f_1 \Delta_{r}}{L}+
\frac{f_2+f_3\Delta_{r}^2}{L^2}+O\left(1/L^3\right)\right]g^{l}=0.
\end{equation}

We next expand $g^l$ in the form
\begin{equation}\label{eq5-90}
g^l =\sum_{k=0}^{\infty}L^{-k}g_k (\Delta_{r},\Delta_{r'},z),
\end{equation}
where the expansion coefficients $g_k$ are considered as being dependent
of only the ``neutral'' variables $\Delta_{r}$,
\begin{equation}\label{eq5-100}
\Delta_{r'}\equiv 2V_{00}^{1/2}L(r'_*-r_{*0}),
\end{equation}
and
\begin{equation}\label{eq5-110}
z\equiv 2\sqrt{xy}=(L/r_0)s.
\end{equation}
Here, $s$ is the geodesic distance, to leading order in $r-r'$,
between the Green's function evaluation and source points
(when these two points have the same $\theta$ and $\varphi$ values):
\begin{equation}\label{eq5-115}
s=\left[f(r_0)(v-v')(u-u')\right]^{1/2}.
\end{equation}
Substituting now the expansion (\ref{eq5-90}) into Eq.\ (\ref{eq5-80})
and comparing powers of $L$, we obtain a hierarchy of
equations for the various functions $g_k$, having the form
\begin{equation}\label{eq5-120}
g_{k,yx}+g_k =S_k.
\end{equation}
Here, the source $S_k$ is determined for each $k>0$ by the functions
$g_{k'<k}$ preceding $g_k$ in the hierarchy.
In the analysis below we shall only need the terms with $k=0$, $1$, and $2$.
For these values of $k$, the source terms are given by
\begin{mathletters}\label{eq5-130}
\begin{equation}\label{eq5-130A}
S_0=0,
\end{equation}
\begin{equation}\label{eq5-130B}
S_1=-f_1\Delta_{r}\,g_0,
\end{equation}
\begin{equation}\label{eq5-130C}
S_2=-f_1\Delta_{r}\,g_1 -(f_2+f_3\Delta_{r}^2)\,g_0.
\end{equation}
\end{mathletters}

Finally, to complete the formulation of a characteristic initial data
problem for each of the functions $g_k$, we supplement Eq.\ (\ref{eq5-120})
with the initial conditions
\begin{equation}\label{eq5-140}
g_k (v=v')=g_k (u=u')=\delta_{k0},
\end{equation}
which conform with the original initial conditions for $g^l$,
Eq.\ (\ref{eq3-120}).

\subsection{Analytic solutions for $k=0$, $1$, and $2$}

The solution to Eq.\ (\ref{eq5-120}) for $k=0$, subject to the initial
conditions, Eq.\ (\ref{eq5-140}), is given by
\begin{equation}\label{eq5-150}
g_0=J_0(z),
\end{equation}
where $J_n$ are the Bessel functions of the first kind, of order $n$.

To solve for $g_1$, we first express the source $S_1$ explicitly as a
function of $y-x$ and $z$, using Eq.\ (\ref{eq5-150}) and the relation
$\Delta_r=y-x+\Delta_{r'}$. We find
\begin{equation}\label{eq5-160}
S_1= -f_1J_0(z)(y-x)-f_1\Delta_{r'}J_0(z).
\end{equation}
Then, with the help of Table \ref{tableI}, we find the solution
for $k=1$ [satisfying Eq.\ (\ref{eq5-140})] to read
\begin{equation}\label{eq5-170}
g_1 =-\frac{1}{4}f_1 zJ_1(z)(\Delta_{r}+\Delta_{r'}).
\end{equation}

We now use the above solutions for $g_0$ and $g_1$ to express $S_2$ as
\begin{equation}\label{eq5-180}
S_2=\frac{1}{4}f_1^2zJ_1(z)\left[(y-x)^2+3\Delta_{r'}(y-x)
+2\Delta_{r'}^2\right]
-f_3J_0(z)\left[(y-x)^2+2\Delta_{r'}(y-x)+\Delta_{r'}^2\right]
-f_2J_0(z).
\end{equation}
With the help of Table \ref{tableI}, we then construct the following
solution for $g_2$, satisfying Eq.\ (\ref{eq5-140}):
\begin{equation}\label{eq5-190}
g_2 =-\frac{1}{6}zJ_1(z)\left[f_3
(\Delta_{r}^2+\Delta_{r}\Delta_{r'}+\Delta_{r'}^2)
                     +3f_2\right]
+\frac{1}{96}z^2J_2(z)\left[3f_1^2(\Delta_{r}+
               \Delta_{r'})^2-8f_3\right]
+\frac{1}{96}f_1^2 z^3J_3(z).
\end{equation}

\begin{table}[h]
\centerline{$\begin{array}{|l|l|} \hline
\hspace{3mm}S(z;x-y) & \text{solution to\ } g_{,yx}+g=S(z;x-y) \\ \hline
J_0(z)               & z J_1(z)/2       \\
(y-x) J_0(z)         & (y-x)zJ_1(z)/4   \\
(y-x)^2 J_0(z)       & \left[z^2J_2(z)+2(y-x)^2zJ_1(z)\right]/12 \\
z J_1(z)             & z^2J_2(z)/4      \\
(y-x) z J_1(z)       & (y-x)z^2J_2(z)/6 \\
(y-x)^2 z J_1(z)     & \left[z^3J_3(z)+3(y-x)^2z^2J_2(z)\right]/24\\
\hline
\end{array}$
}
\caption{\protect\footnotesize
Specific solutions to the inhomogeneous partial DE $g_{,yx}+g=S$
for various source functions of the form $S(z;x-y)$, where
$z=2\sqrt{xy}$. These solutions can be verified by substitution.
In this table, $J_n$ are the Bessel functions of the first kind,
of order $n$.}
\label{tableI}
\end{table}

\subsection{$G^{l\pm}_{,\alpha}$ expanded in powers of $1/L$}

We are now in position to write the three leading-order
terms in the $1/L$ expansion of th egradient
$G^{l\pm}_{,\alpha}[x_0^{\mu};x_p^{\mu}]$.
To that end we shall need, in view of Eqs.\ (\ref{eq4-100}) and
(\ref{eq4-110}), to evaluate the functions $g_k$ derived above, along with
their gradients $g_{k,\alpha}$, for $x^{\mu}=x_0^{\mu}$ and
${x'}^{\mu}=x_p^{\mu}$.
For the calculation of $g_{k,\alpha}$ it is convenient to
use the auxiliary relations
\begin{equation}\label{eq5-200}
d[z^n J_n(z)]/dz=z^nJ_{n-1}(z)
\end{equation}
[for $n=0$ recall $J_{-1}(z)=-J_1(z)$], along with
\begin{equation}\label{eq5-220}
dz/dr=-f_0 L(y-x)/z \quad\text{and}\quad
dz/dt= f f_0 L(y+x)/z,
\end{equation}
where
\begin{equation}\label{eq5-270}
f_0\equiv [r_0f^{1/2}(r_0)]^{-1}.
\end{equation}
With the help of these relations we derive from Eqs.\ (\ref{eq5-150}),
(\ref{eq5-170}), and (\ref{eq5-190}) expressions for $g_{k,r}$ and
$g_{k,t}$ (where $k=0,1,2$). We then set $x^{\mu}=x_0^{\mu}$ and
${x'}^{\mu}=x_p^{\mu}$ in these expressions, and also in the
expressions for the functions $g_k$ themselves (noticing the
vanishing of $\Delta_r$).
All resulting expressions are then substituted in the formulas
for $G^{l\pm}_{,r}$ and $G^{l\pm}_{,t}$, Eqs.\ (\ref{eq4-100})
and (\ref{eq4-110}). In these equations we also make the substitution
$\delta(w_{\pm}-w'_{\pm})=2LV_{00}^{1/2}\delta(\hat w_{\pm})$
where $\hat w_{\pm}$ are `neutral' variables defined by
\begin{equation}\label{eq5-225}
\hat w_{+}\equiv 2x \quad\text{and}\quad \hat w_{-}\equiv 2y.
\end{equation}
Finally, collecting common powers of $L$ we obtain, for $\alpha=r$ or $t$,
an expression of the form
\begin{equation}\label{eq5-230}
G^{l\pm}_{,\alpha}[x_0^{\mu};x_p^{\mu}]=
\frac{P_{l}(\cos\chi)}{r r'}\left(
\hat H_{\alpha}^{(0)\pm}L^2 + \hat H_{\alpha}^{(1)}L +
\hat H_{\alpha}^{(2)} +\cdots\right),
\end{equation}
where the various coefficients $\hat H_{\alpha}$ are functions of only the
`neutral' spacetime-interval variables\footnote{We hereafter use the symbols
$\Delta_{r'}$, $z$, and $\hat w_{\pm}$ to represent the values of these
variables for $x^{\mu}=x_0^{\mu}$ and ${x'}^{\mu}=x_p^{\mu}$.}
$\Delta_{r'}$, $z$, $\hat w_{\pm}$, and
\begin{equation}\label{eq5-240}
\Delta_t\equiv 2V_{00}^{1/2}L(t_0-t_p).
\end{equation}
These coefficient functions are given by
\begin{mathletters}\label{eq5-250}
\begin{equation}\label{eq5-250A}
\hat H_r^{(0)\pm}=f_0\left[-\Delta_{r'}J_1(z)/z
                    \mp J_0(z)\delta(\hat w_{\pm})\right],
\end{equation}
\begin{equation}\label{eq5-250B}
\hat H_r^{(1)}=-\frac{1}{4}f_0f_1
    \left[zJ_1(z) + \Delta_{r'}^2 J_0(z)\right] - J_0(z)/r_0,
\end{equation}
\begin{equation}\label{eq5-250C}
\hat H_r^{(2)}=\frac{1}{96}f_0 \Delta_{r'}\left[
    7f_1^2z^2J_2(z)+3zJ_1(z)(f_1^2 \Delta_{r'}^2 -8f_3)-
    16J_0(z)(f_3\Delta_{r'}^2+3f_2)\right]
    +\frac{1}{4}f_1 \Delta_{r'}zJ_1(z)/r_0,
\end{equation}
\end{mathletters}
and
\begin{mathletters}\label{eq5-260}
\begin{equation}\label{eq5-260A}
\hat H_t^{(0)\pm}=ff_0\left[-\Delta_{t} J_1(z)/z
                    + J_0(z)\delta(\hat w_{\pm})\right],
\end{equation}
\begin{equation}\label{eq5-260B}
\hat H_t^{(1)}=-\frac{1}{4}ff_0f_1 \Delta_{r'}\Delta_t J_0(z),
\end{equation}
\begin{equation}\label{eq5-260C}
\hat H_t^{(2)}=\frac{1}{96}ff_0 \Delta_{t}\left[
    f_1^2z^2J_2(z)+zJ_1(z)(3f_1^2 \Delta_{r'}^2 -8f_3)-
    16J_0(z)(f_3\Delta_{r'}^2+3f_2)\right]
\end{equation}
\end{mathletters}
(where the function $f$ is to be evaluated at $r=r_0$).
In the above expressions for $\hat H_r^{(1),(2)}$ and
$\hat H_t^{(1),(2)}$ we have omitted terms of the form
$\propto z^k J_n(z)\delta(\hat w_{\pm})$, with $k$ being a positive
integer, as such terms would yield vanishing contributions to
$\delta F_{\alpha}^{l(\epsilon)\pm}$ when integrated over
$\tau$ in Eq.\ (\ref{eq4-90}).

The quantities $\delta F^{l(\epsilon)\pm}_{\alpha}$ can now be
constructed, in principle, for any given worldline, by inserting
Eq.\ (\ref{eq5-230}) into Eq.\ (\ref{eq4-90}) and carrying out the
integration over $\tau$.
In practice, to integrate over $\tau$, one should proceed as follows:
Recalling that $\tau$ is a small quantity (we have $|\tau|\leq\epsilon$),
one first expands in powers of $\tau$ all $\tau$-dependent quantities in
the integrand of Eq.\ (\ref{eq4-90}).
(At that point, the details of the specific trajectory
under consideration enter the calculation in an explicit manner;
specifically, the power expansion coefficients turn out to depend on
the values of $u^{\alpha}$, $\dot u^{\alpha}$, and $\ddot u^{\alpha}$
at the force's evaluation point.)
Then, since we are interested in extracting the large $l$
(large $L$) behavior of $\delta F^{l(\epsilon)\pm}_{\alpha}$, we
introduce the ``neutral'' dimensionless proper time variable, defined by
\begin{equation}\label{eq5-280}
\lambda\equiv -(L/r_0)\tau,
\end{equation}
and replace $\tau$ by $-r_0(\lambda/L)$.
The integrand then takes the form of a power series in $1/L$,
with $\lambda$-dependent coefficients.
Transforming finally from integration over $\tau$ to integration over
$\lambda$, one obtains an expression for $\delta F^{l(\epsilon)\pm}_
{\alpha}$ in the form of a power series in $1/L$, as desired.
In the rest of this paper we carry out the above calculation in
full detail (and derive $h^{l}_{\alpha}$ and $D_{\alpha})$ for the
case of a purely radial trajectory.

\section{Form of the regularization function $\lowercase{h}^
    {\lowercase{l}}$: the case of radial motion}\label{secVI}

When considering radial trajectories (namely, ones along which
$d\theta=d\varphi=0$) one has $P_{l}(\cos\chi)\equiv 1$.
Consequently, the Green's function given in Eq.\ (\ref{eq3-150})
becomes $\theta,\varphi$-independent, resulting in the vanishing
of both angular components of the self force (as should be expected,
of course, by virtue of the background being spherically-symmetric).
In the following we discuss the $r$ and $t$ components
of the self-force.

To carry out the integration in Eq.\ (\ref{eq4-90}) we
first expand each $\tau$-dependent quantity in the integrand in powers
of $1/L$, with $\lambda$ held fixed. The $\tau$-dependent quantities
to be expanded are $\Delta_{r'}$, $\Delta_t$, $1/r'$, $z$, and the various
Bessel functions appearing in Eqs.\ (\ref{eq5-250}) and (\ref{eq5-260}).

By expanding $\Delta_{r'}$ in a Taylor series in $\tau$ about $r=r_0$, and
transforming to the variable $\lambda$, we obtain the expansion
\begin{eqnarray}\label{eq6-10}
\Delta_{r'}&=&\dot{\Delta}_{r'}\tau+\frac{1}{2}\ddot{\Delta}_{r'}\tau^2+
\frac{1}{6}\overdots{\Delta}_{r'}\tau^3+\cdots=\nonumber\\
&& f^{1/2}\left[-\dot{r}_*\lambda+\frac{1}{2}\ddot{r}_*\lambda^2(r_0/L)-
\frac{1}{6}\overdots{r}_*\lambda^3(r_0/L)^2\right]+O(1/L)^3,
\end{eqnarray}
where
\begin{eqnarray}\label{eq6-20}
\dot{r}_*&=&  f^{-1} \dot{r},                 \nonumber\\
\ddot{r}_*&=& f^{-2}(f\ddot{r}-f'\dot{r}^2),  \nonumber\\
\overdots{r}_*&=& f^{-3}\left[(2f'^2-f''f)\dot{r}^3-
3f'f\dot{r}\ddot{r}+f^2\overdots{r}\right],
\end{eqnarray}
and where all quantities (except $\lambda$) are evaluated at $r=r_0$
($\tau=0$). In a similar manner we obtain for $\Delta_t$,
\begin{equation}\label{eq6-25}
\Delta_t=
f^{1/2}\left[\dot{t}\lambda-\frac{1}{2}\ddot{t}\lambda^2 (r_0/L)+
\frac{1}{6}\overdots{t}\lambda^3 (r_0/L)^2\right]+O(1/L)^3,
\end{equation}
and for $1/r'$,
\begin{equation}\label{eq6-30}
\frac{1}{r'}=\frac{1}{r_0}\left[1+\dot{r}(\lambda/L)
+\frac{1}{2}(2\dot{r}^2-r_0\ddot{r})(\lambda/L)^2)\right]+
O(1/L)^3.
\end{equation}

Next, recalling $z=(L/r_0)s$, we obtain
\begin{equation}\label{eq6-40}
z=
-\dot{s}\lambda+\frac{1}{2}\ddot{s}\lambda^2 (r_0/L)-
\frac{1}{6}\overdots{s}\lambda^3 (r_0/L)^2+O(1/L)^3.
\end{equation}
To calculate the $\tau$-derivatives of $s$
(which are understood here to be evaluated at $\tau=0$),
we make use of the normalization relation $\dot{v}\dot{u}=1/f$,
and of the relations derived from it by successively differentiating
its both sides with respect to $\tau$:
$\ddot{v}\dot{u}+\dot{v}\ddot{u}=(1/f)'\dot{r}$, and
$\overdots{v}\dot{u}+2\ddot{v}\ddot{u}+\dot{v}\overdots{u}=
(1/f)''\dot{r}^2+(1/f)'\ddot{r}$.
Using these relations we find
\begin{eqnarray}\label{eq6-60}
\dot{s}(\tau=0)&=&-1,
\\              \label{eq6-70}
\ddot{s}(\tau=0)&=&\frac{1}{2}(f'/f)\dot{r},
\\              \label{eq6-80}
\overdots{s}(\tau=0)&=&
\frac{1}{16f^2}\left[\left(8f''f-13{f'}^2\right)\dot{r}^2
+8f'f\ddot{r}\right] + \frac{1}{4}f \ddot{v}\ddot{u}
\end{eqnarray}
(note that whereas $\tau$ is
non-positive throughout the integration domain, $z$ and $s$ are, by
definition, non-negative).

Finally, we need to similarly expand the various Bessel functions appearing
in the integrand of Eq.\ (\ref{eq4-90}).
Using Eqs.\ (\ref{eq6-40}) and (\ref{eq6-60}) we find for any $n\geq 0$,
\begin{eqnarray}\label{eq6-90}
J_n(z)=
J_n(\lambda)+\frac{1}{2}(r_0/L)\ddot{s}\lambda^2J'_n(\lambda)+
(r_0/L)^2\left(\frac{1}{8}\ddot{s}^2\lambda^4 J''_n(\lambda)
-\frac{1}{6} \overdots{s}\lambda^3J'_n(\lambda)
\right) +O(1/L^3),
\end{eqnarray}
where a prime denotes $d/d\lambda$.
Using this general form together with Eq.\ (\ref{eq5-200}), we obtain
the following expansions, needed for our analysis:
\begin{eqnarray}\label{eq6-100}
J_0(z)&=&J_0(\lambda)-\frac{1}{2}(r_0/L)\ddot{s}\lambda^2
J_1(\lambda)+ O(1/L^2)
\\ \label{eq6-105}
J_1(z)&=& J_1(\lambda)+
\frac{1}{2}(r_0/L)\ddot{s}
\left(\lambda^2J_0(\lambda)-\lambda J_1(\lambda)\right)
\nonumber\\  &&
+(r_0/L)^2\left[\frac{1}{8}\ddot{s}^2
\left(\lambda^3J_2(\lambda)-\lambda^4J_1(\lambda)\right)
-\frac{1}{6}\overdots{s}\left(\lambda^3J_0(\lambda)-
\lambda^2J_1(\lambda)\right)\right]
+ O(1/L^3).
\end{eqnarray}

We now substitute the above expansions for $\Delta_{r'}$, $\Delta_t$,
$z$, and the Bessel functions in Eqs.\ (\ref{eq5-250})--(\ref{eq5-260}).
We also substitute for the delta functions in Eqs.\ (\ref{eq5-250A})
and (\ref{eq5-260A})
\begin{equation}\label{eq6-110}
\delta(\hat w_{\pm})=
\frac{\delta(\lambda)}{\left|d\hat w_{\pm}/d\lambda\right|}=
\frac{\delta(\lambda)}{f^{1/2}\dot{w}_{\pm}}=
f^{1/2}\dot{w}_{\mp}\delta(\lambda),
\end{equation}
where the last equality is due to the normalization of the four-velocity.
We thereby obtain expressions for the various functions
$\hat H_{\alpha}^{(n=0,1,2)}$, each expanded in powers of
$1/L$ up to order $O(L^{n-2})$ (with $\lambda$-dependent coefficients).
Substitution of these expressions [and of the expansion for $1/r'$, Eq.\
(\ref{eq6-30})] into Eq.\ (\ref{eq5-230}) finally yields the desired
expression for the Green's function's gradient, as a power series in
$1/L$ (with $\lambda$ held fixed).
We find (for $\alpha=r,t$)
\begin{equation}\label{eq6-120}
G_{,\alpha}^{l\pm}=H_{\alpha}^{(0)\pm}L^2 + H_{\alpha}^{(1)}L +
H_{\alpha}^{(2)} + O(1/L),
\end{equation}
where the various coefficients $H_{\alpha}^{(n)}$
are functions of $\lambda$ along the worldline, given by
\begin{equation}\label{eq6-130}
H_r^{(0)\pm}= \frac{1}{r_0^3}\left[\dot{r}_* J_1(\lambda)
\mp \dot{w}_{\mp} J_0(\lambda)\delta(\lambda)\right],
\end{equation}
\begin{eqnarray}\label{eq6-140}
H_r^{(1)}= -\frac{1}{4r_0^3}\left[\lambda J_1(\lambda)
\left(f_1/f^{1/2}-4\dot{r}\dot{r}_*+4r_0\ddot{s}\dot{r}_*
+2r_0\ddot{r}_*\right)\right. \nonumber\\
+\left.\lambda^2 J_0(\lambda)\left(f^{1/2}f_1\dot{r}_*^2-
2r_0\ddot{s}\dot{r}_* \right)+ 4J_0(\lambda)\right],
\end{eqnarray}
\begin{eqnarray}\label{eq6-150}
H_r^{(2)} &=& -\frac{1}{96 r_0^3}\left[
    48\lambda J_0(\lambda)\left(2\dot{r}-f_2 \dot{r}_*\right)
    \right.\nonumber\\&&
    +4\lambda^3 J_0(\lambda)\left(
    3f_1r_0\ddot{s}/f^{1/2}+6r_0^2\ddot{s}^2\dot{r}_*
    +4r_0^2\overdots{s}\dot{r}_*-4ff_3\dot{r}_*^3+
    6r_0^2\ddot{s}\ddot{r}_*-6f^{1/2}f_1r_0\dot{r}_*\ddot{r}_*
    \right.\nonumber\\&& \hspace{20mm}\left.
    -12\dot{r}r_0\ddot{s}\dot{r}_* +6f^{1/2}f_1\dot{r}\dot{r}_*^2
    \right)\nonumber\\ &&
    +8\lambda^2 J_1(\lambda)\left(
    3f_1 \dot{r}/f^{1/2}-6r_0\ddot{s}-3f_3\dot{r}_*
    -12\dot{r}^2\dot{r}_*+6r_0\ddot{r}\dot{r}_*
    +12r_0\dot{r}\ddot{s}\dot{r}_*
    \right.\nonumber\\&& \hspace{20mm}\left.
    -6r_0^2\ddot{s}^2\dot{r}_*
    -4r_0^2\overdots{s}\dot{r}_*+6r_0\dot{r}\ddot{r}_*
    -6r_0^2\ddot{s}\ddot{r}_*-2r_0^2\overdots{r}_*
    +3f^{1/2}f_1\dot{r}_*
    \right)\nonumber\\ &&
    +3\lambda^4 J_1(\lambda)\left(
    4r_0^2\ddot{s}^2\dot{r}_*+ff_1^2\dot{r}_*^3
    -4f^{1/2}f_1r_0\ddot{s}\dot{r}_*^2
    \right)\nonumber\\ &&
    \left.
    +\lambda^3 J_2(\lambda)\left(
    7f_1^2\dot{r}_*-12r_0^2\ddot{s}^2\dot{r}_*\right)\right],
\end{eqnarray}
and
\begin{equation}\label{eq6-160}
H_t^{(0)\pm}= \frac{f}{r_0^3}\left[J_0(\lambda)\dot{w}_{\mp}
\delta(\lambda)- J_1(\lambda) \dot{t}\right],
\end{equation}
\begin{equation}\label{eq6-170}
H_t^{(1)}= \frac{f}{4r_0^3}\left\{\lambda J_1(\lambda)
\left[-4\dot{r}\dot{t}+2r_0\left(2\ddot{s}\dot{t}+\ddot{t}\right)\right]
+\lambda^2 J_0(\lambda)\left(-2r_0\ddot{s}\dot{t}+
f^{1/2}f_1\dot{r}_*\dot{t}\right)\right\},
\end{equation}
\begin{eqnarray}\label{eq6-180}
H_t^{(2)} &=& -\frac{f}{96 r_0^3}\left\{
    48\lambda J_0(\lambda)f_2\dot{t} \right.\nonumber\\&&
    -4\lambda^3 J_0(\lambda)\left[
                r_0^2\left(6\ddot{s}^2\dot{t}+4\overdots{s}\dot{t}+6\ddot{s}\ddot{t}
                \right)-3r_0\left(4\dot{r}\ddot{s}\dot{t}+f^{1/2}f_1\ddot{t}
                \dot{r}_*+f^{1/2}f_1\dot{t}\ddot{r}_*\right)+6f^{1/2}f_1\dot{r}
                \dot{t}\dot{r}_*-4ff_3\dot{t}\dot{r}_*^2\right]
                \nonumber\\ &&
    +8\lambda^2 J_1(\lambda)\left(
                f_3\dot{t}+12\dot{r}^2\dot{t}-6r_0\ddot{r}\dot{t}-12r_0\dot{r}
                    \ddot{s}\dot{t}+6r_0^2\ddot{s}^2\dot{t}+4r_0^2\overdots{s}\dot{t}
                    -6r_0\dot{r}\ddot{t}+6r_0^2\ddot{s}\ddot{t}+2r_0^2\overdots{t}
                    \right)  \nonumber\\ &&
    +3\lambda^4 J_1(\lambda)\left(
    -4r_0^2\ddot{s}^2\dot{t}+4f^{1/2}f_1r_0\ddot{s}\dot{t}\dot{r}_*
    -ff_1^2\dot{t}\dot{r}_*^2
    \right)\nonumber\\ &&
    \left. +\lambda^3 J_2(\lambda)
                \left(-f_1^2\dot{t}+12r_0^2\ddot{s}^2\dot{t}\right)\right\}.
\end{eqnarray}

Changing the integration variable in Eq.\ (\ref{eq4-90}) from $\tau$
to $\lambda$, we now have
\begin{equation}\label{eq6-190}
\delta F_{\alpha}^{l(\epsilon)\pm}=q^2 r_0\int_{0}^{L\epsilon/r_0}
\left[LH_{\alpha}^{(0)\pm}+H_{\alpha}^{(1)}+H_{\alpha}^{(2)}/L+
O(L^{-2})\right] d\lambda.
\end{equation}
The desired regularization function $h^{l}_{\alpha}$ is to be constructed
such as to extract the large $l$ singular behavior of
$\delta F_{\alpha}^{l(\epsilon)\pm}$ (while maintaining the simplest form
possible). By virtue of Eq.\ (\ref{eq6-190}) we take this function
to be
\begin{equation}\label{eq6-200}
h^{l\pm}_{\alpha}=LA_{\alpha}^{\pm}+B_{\alpha}+C_{\alpha}/L,
\end{equation}
where
\begin{mathletters} \label{eq6-210}
\begin{eqnarray}\label{eq6-210A}
A_{\alpha}^{\pm}&\equiv&\widetilde\lim_{l\to\infty}
        \left(L^{-1}\,\delta F_{\alpha}^{l(\epsilon)\pm}\right)
        =q^2 r_0 \int_{0}^{\widetilde{\infty}}
        H_{\alpha}^{(0)\pm}(\lambda)d\lambda,
\end{eqnarray}
\begin{eqnarray}\label{eq6-210B}
B_{\alpha}&\equiv&\widetilde\lim_{l\to\infty}
        \left(\delta F_{\alpha}^{l(\epsilon)\pm}-LA_{\alpha}^{\pm}\right)
         =q^2 r_0 \int_{0}^{\widetilde\infty}
         H_{\alpha}^{(1)}(\lambda)d\lambda,
\end{eqnarray}
\begin{eqnarray}\label{eq6-210C}
C_{\alpha}&\equiv&\widetilde\lim_{l\to\infty}
        L\left(\delta F_{\alpha}^{l(\epsilon)\pm}-LA_{\alpha}^{\pm}-
        B_{\alpha}\right)
        = q^2 r_0\int_{0}^{\widetilde\infty}
        H_{\alpha}^{(2)}(\lambda)d\lambda,
\end{eqnarray}
\end{mathletters}
with $\int^{\widetilde\infty}(\ )d\lambda$ standing for
$\widetilde\lim_{x\to\infty}\int^x(\ )d\lambda$.\footnote
{Here we extend the definition of the tilde-limit, given in App.\
\ref{AppA}, from discrete functions with index $l$ to continuous functions
of $\lambda$. In an analogous manner, the tilde-limit of a function
$f(\lambda)$ as $\lambda\to\infty$ would be defined through the subtraction
of a finite sum of functions of the form
$B^{(j)}(\lambda)=a_j \lambda^{b_j}\cos(\alpha_j \lambda+\beta_j)$
(with $\alpha_j\neq 0$ for all $j$).
It is simple to verify that the tilde-limit of a function, when existing,
is single-valued.}
That the second equality in each of Eqs.\ (\ref{eq6-210B}) and
(\ref{eq6-210C}) is valid, and that the above choice of function
$h^{l\pm}_{\alpha}$ indeed satisfies the requirement that the
tilde-sum $\sum_{l=0}^{\widetilde\infty} \left(\delta
F^{l(\epsilon)\pm}_{\alpha}-h^{l\pm}_{\alpha}\right)$
would converge, will be shown in the next section,
where we explicitly calculate the parameters $A_{\alpha}$, $B_{\alpha}$,
and $C_{\alpha}$. The reason for using the tilde-limit instead of the
standard limit in the definitions of these parameters will also become
clear then. As to the parameter $D_{\alpha}$, substituting now Eqs.\
(\ref{eq6-190}) and (\ref{eq6-200}) into Eq.\ (\ref{eq4-70}), one
obtains\footnote{
We assume here that the contribution associated with the $O(L^{-2})$ term
in Eq.\ (\ref{eq6-190}) vanishes upon taking the limit $\epsilon \to 0$,
as this contribution is of order $O(\epsilon)$ (this becomes clear from
the calculation of $D_{\alpha}$ in the next section). However,
a problem may occur if this term fails to yield a finite contribution
when integrated over $\lambda$. Here we do not further investigate
the behavior of the $O(L^{-2})$ term, and just assume that the above
potential problem is not realized.
}
\begin{equation}\label{eq6-220}
D_{\alpha}=-q^2 r_0 \lim_{\epsilon\to 0}\sum_{l=0}^{\widetilde\infty}
\int_{L\epsilon/r_0}^{\widetilde\infty} \left[LH_{\alpha}^{(0)\pm}+
H_{\alpha}^{(1)}+H_{\alpha}^{(2)}/L \right] d\lambda.
\end{equation}

In conclusion, we find the tail part of the self force to be given by
\begin{equation}\label{eq6-230}
F^{\rm (tail)}_{\alpha}=\sum_{l=0}^{\infty}
\left(F^{l\pm}_{\alpha}- A_{\alpha}^{\pm}L-B_{\alpha}-
C_{\alpha}/L\right)-D_{\alpha},
\end{equation}
where, from the above construction of $h_{\alpha}^{l}$, it follows that
the sum over $l$ converges at least as $\sim 1/l$. The
implementation of our regularization scheme thus amounts to analytically
determining the regularization parameters $A_{\alpha}^{\pm}$, $B_{\alpha}$,
$C_{\alpha}$, and $D_{\alpha}$, using Eqs.\ (\ref{eq6-210}) and
(\ref{eq6-220}). For the calculation of the tail term, one may use
either $F^{l+}_{\alpha}$ (with $A_{\alpha}^{+}$) or $F^{l-}_{\alpha}$
(with $A_{\alpha}^{-}$). Of course, one may also use any combination of
these two one-sided quantities (e.g., their average). It should be emphasized
here that the {\em final} result of the calculation, namely the tail term
$F^{\rm(tail)}_{\alpha}$ (having a well defined value at the evaluation
point), should be the same regardless of whether it is derived from one of
the one-sided limits, or from the other, or, say, from their average.

We finally point out that, although Eq.\ (\ref{eq6-230}) has been
developed here for radial motion, an expression of this form is also
valid for any other trajectory \cite{unpublished}. The details of the
specific trajectory under consideration would only affect the values
of the various regularization parameters.

\section{Derivation of the regularization parameters for radial motion}
\label{secVII}

To carry out the calculation of the regularization parameters in this
section, we shall need the following integrals, the derivation of which
will be described in Appendix \ref{AppB}.
For $k,n\in\mathbb{N}$ we have
\begin{equation}\label{eq7-5}
\int_0^{\widetilde\infty}\lambda^k J_n(\lambda)d\lambda=\left\{
\begin{array}{ll}
(n+k-1)!!/(n-k-1)!!,                 & 0\leq k\leq n,        \\
(-1)^{(k-n)/2}(k+n-1)!!(k-n-1)!!,    & \text{even $k-n>0$},  \\
0,                                   & \text{odd  $k-n>0$}
\end{array}\right.
\end{equation}
[in applying this formula for $k=n$, recall $(-1)!!=1$].
If the difference $k-n$ is a positive odd integer, then we also have
\begin{equation}\label{eq7-6}
\int_0^{\widetilde\infty}\lambda^k J_n(\lambda)\ln\lambda \,d\lambda=
(-1)^{(k-n+1)/2}(k+n-1)!!(k-n-1)!!.
\end{equation}

\subsection{Derivation of $A_{\alpha}^{\pm}$}

Substituting Eqs.\ (\ref{eq6-130}) and (\ref{eq6-160}) into Eq.\
(\ref{eq6-210A}) and carrying out the tilde-integration [with the help of
Eq.\ (\ref{eq7-5})] we find, recalling $J_0(0)=1$ and $J_0(\infty)=0$,
\begin{eqnarray}\label{eq7-10}
A_{r}^{\pm}&=&
\frac{q^2}{r_0^2}\left(\dot{r}_* \mp \dot{w}_{\mp}\right)=
\mp\frac{q^2}{r_0^2}\,\dot{t},\nonumber\\
A_{t}^{\pm}&=&
f\frac{q^2}{r_0^2}\left(-\dot{t} + \dot{w}_{\mp}\right)=
\pm\frac{q^2}{r_0^2}\,\dot{r}.
\end{eqnarray}
Note that the two one-sided values of $A_{\alpha}$ are, in general,
not the same.
Consequently, as argued above, the function $h_{\alpha}^{l}$ (and
also $\delta F_{\alpha}^{l(\epsilon)}$) exhibits two different
one-sided values.
We note that the {\em averaged} value of the parameter $A_{\alpha}$,
to be denoted by $\bar{A}_{\alpha}$, is found to vanish:
\begin{equation}\label{eq7-20}
\bar{A}_{\alpha} \equiv \frac{1}{2}(A_{\alpha}^++A_{\alpha}^-)=0.
\end{equation}
This vanishing of $\bar{A}_{\alpha}$ seems to occur for all trajectories
of a scalar particle, not only the radial ones considered here
\cite{unpublished}.

\subsection{Derivation of $B_{\alpha}$}

By the definition of $B_{\alpha}$ in Eq.\ (\ref{eq6-210B}) we have,
after substituting for $A_{\alpha}^{\pm}$ from Eq.\ (\ref{eq6-210A}),
\begin{equation}\label{eq7-25}
B_{\alpha}=-q^2 r_0 \,\widetilde\lim_{l\to\infty}\left[
L\int_{L\epsilon/r_0}^{\widetilde\infty}H_{\alpha}^{(0)\pm}(\lambda)\,
d\lambda \right]
+q^2 r_0 \,\int_{0}^{\widetilde\infty} H_{\alpha}^{(1)}(\lambda)
\,d\lambda.
\end{equation}
The first term here cancels out upon taking the tilde-limit
$l\to\infty$: At large $l$ (and fixed $\epsilon$), each of the
components $H_{\alpha}^{(0)\pm}$ behaves as $\propto l^{-1/2}$ times
oscillations with respect to $l$ [these components are linear
combinations of Bessel functions, the asymptotic form of which is
described in Eq.\ (\ref{eqB-200})]. To leading order in $1/l$, this
is also the form of the integral over $H_{\alpha}^{(0)\pm}$ in Eq.\
(\ref{eq7-25}) (which is carried out over asymptotically large
values of $\lambda$).
Hence, the expression in the squared brackets is found to diverge
as $\propto l^{1/2}$ times oscillations.
When taking the tilde-limit, this divergent piece is removed,
with the remaining part dying off at large $l$ as $\propto l^{-1/2}$
(times oscillations). Therefore, no contribution arises from the
first term in Eq.\ (\ref{eq7-25}), and the second equality of
Eq.\ (\ref{eq6-210B}) is shown to be valid.

To calculate the parameter $B_{\alpha}$, we now substitute for
$H_{\alpha}^{(1)}$ from Eqs.\ (\ref{eq6-140}) and (\ref{eq6-170})
(for the $r$ and $t$ components, respectively).
The calculation involves tilde-integrating over terms of the form
$\propto\lambda J_1(\lambda)$, $\propto\lambda^2 J_0(\lambda)$, and
$\propto J_0(\lambda)$. Reading the values of the these integrals
from Eq.\ (\ref{eq7-5}), and substituting for $f_1$, $\dot{r}_*$,
$\ddot{r}_*$, and $\ddot{s}$ [using Eqs.\ (\ref{eq5-60}),
(\ref{eq6-20}), and (\ref{eq6-70})], we obtain
\begin{equation}\label{eq7-50}
B_r=-\frac{q^2}{2r_0^2}f^{-1}\left(f+r_0f'/2-\dot{r}^2
  +r_0\ddot{r}\right)
\end{equation}
and
\begin{equation}\label{eq7-60}
B_t=\frac{q^2}{2r_0^2}\left[fr_0\ddot{t}+\dot{t}\dot{r}(r_0f'-f)\right].
\end{equation}
Recalling $a^r=\ddot{r}+f'/2$ and $a^t=\ddot{t}+f'\dot{r}_*\dot{t}$
in the case of radial motion considered here,
we may express this result in a more compact form as
\begin{equation}\label{eq7-70}
B_{\alpha}=-\frac{q^2}{2r_0^2}\left(\delta_{\alpha}^{r}+
r_0 a_{\alpha} - \dot r u_{\alpha}\right).
\end{equation}

\subsection{Derivation of $C_{\alpha}$}

By its definition in Eq.\ (\ref{eq6-210C}), we have for the
parameter $C_{\alpha}$, after substituting for $A_{\alpha}$ and
$B_{\alpha}$,
\begin{equation}\label{eq7-75}
C_{\alpha}=-q^2 r_0 \,\widetilde\lim_{l\to\infty}\left[
\int_{L\epsilon/r_0}^{\widetilde\infty}
\left(LH_{\alpha}^{(0)}(\lambda)+H_{\alpha}^{(1)}(\lambda)\right)\,
d\lambda \right]
+q^2 r_0 \,\int_{0}^{\widetilde\infty} H_{\alpha}^{(2)}(\lambda)
\,d\lambda.
\end{equation}
Again, there is a residual part left from the calculation of $A_{\alpha}$
and $B_{\alpha}$, which involves integration over asymptotically large
values of $\lambda$. This part can again be shown to vanish as the
tilde-limit $l\to\infty$ is taken, resulting in that only the
second integral in Eq.\ (\ref{eq7-75}) survives.

To calculate $C_{\alpha}$, we thus use the second equality of
Eq.\ (\ref{eq6-210C}), in which we substitute for $H_{\alpha}^{(2)}$
from Eqs.\ (\ref{eq6-150}) and (\ref{eq6-180}) (for the $r$ and $t$
components, respectively).  One then has to evaluates the tilde-integral
of a sum of various terms of the form $\propto\lambda^k J_n(\lambda)$,
all of which have $k>n$ and odd $k-n$. According to Eq.\ (\ref{eq7-5}),
{\em all such integrals vanish}. Thus, we find
\begin{eqnarray}\label{eq7-80}
C_{\alpha}=0.
\end{eqnarray}
The vanishing of the parameter $C_{\alpha}$ seems to be a universal
feature of our scheme, regardless of the specific trajectory under
consideration \cite{unpublished}. As we also find below, this vanishing
constitutes a necessary condition for the self-consistency of the
whole regularization scheme.

\subsection{Derivation of $D_{\alpha}$}

To calculate the parameter $D_{\alpha}$ we write Eq.\ (\ref{eq6-220})
in the form
\begin{equation}\label{eq7-90}
D_{\alpha}=D_{\alpha}^{(0)}+D_{\alpha}^{(1)}+D_{\alpha}^{(2)},
\end{equation}
where
\begin{eqnarray}\label{eq7-100}
D_{\alpha}^{(n)} \equiv -q^2 r_0\lim_{\epsilon\to 0}\sum_{l=0}^
{\widetilde\infty} \int_{L\epsilon/r_0}^{\widetilde\infty}
L^{1-n}H_{\alpha}^{(n)\pm} (\lambda)\,d\lambda.
\end{eqnarray}
In calculating the above three pieces of $D_{\alpha}$, we shall transform
from summation over $l$ to integration over a continuous variable.
For this transformation we will make use of the relation
\begin{equation}\label{EQ7-110}
\sum_{l=0}^{\infty}(\epsilon/r_0)K(L\epsilon/r_0)
=\int_{0}^{\infty}\left[K(x)-\frac{1}{24}(\epsilon/r_0)^2 K''(x)\right]dx
+O(\epsilon^3),
\end{equation}
where $K(x)$ is any (sufficiently regular) integrable real function,
and $x$ is an integration variable.
Here, the $O(\epsilon^0)$ term on the right-hand side (RHS) is the standard
(``Riemann type'') integral, which is obtained, by definition, when the
$\epsilon\to 0$ limit of the left-hand side is taken. We also indicated
here the $O(\epsilon^2)$ correction to the integral, which we shall have
to take into account in the calculation below (it is straightforward
to verify the form of this correction term using standard calculus).
Obviously, Eq.\ (\ref{EQ7-110}) also holds for the tilde-sum
$\sum_{l=0}^{\widetilde\infty}$, where on the RHS we use the
tilde-integral $\int_{0}^{\widetilde\infty}$.

Beginning with the calculation of $D_{\alpha}^{(0)}$, we write Eq.\
(\ref{eq7-100}) for $n=0$ as
\begin{equation}\label{eq7-120}
D_{\alpha}^{(0)}=-q^2 r_0\lim_{\epsilon\to 0}\left[(r_0/\epsilon)^2
\sum_{l=0}^{\widetilde\infty}(\epsilon/r_0)
\left(\left(L\epsilon/r_0\right)
\int_{L\epsilon/r_0}^{\widetilde\infty}
H_{\alpha}^{(0)\pm}(\lambda)\, d\lambda\right)\right].
\end{equation}
Comparing the form of the sum in this expression to the left-hand side
of Eq.\ (\ref{EQ7-110}), we find
\begin{equation}\label{eq7-130}
D_{\alpha}^{(0)}= -q^2 r_0\lim_{\epsilon\to 0}\left[(r_0/\epsilon)^2
\int_{0}^{\widetilde\infty}dx\,\, \left(K_{\alpha}^{(0)}(x)-
\frac{1}{24}(\epsilon/r_0)^2 [K_{\alpha}^{(0)}(x)]''\right)\right],
\end{equation}
where
\begin{equation}\label{eq7-140}
K_{\alpha}^{(0)}(x)\equiv x \int_{x}^{\widetilde\infty} H_{\alpha}^{(0)\pm}
(\lambda)\,d\lambda.
\end{equation}
Note here how the contribution to $D_{\alpha}^{(0)}$ due to the
$O(\epsilon^3)$ term appearing in Eq.\ (\ref{EQ7-110}) vanishes upon
taking the limit $\epsilon\to 0$. We recall that each of the two
components $H_{r}^{(0)\pm}$ and $H_{t}^{(0)\pm}$, given explicitly
in Eqs.\ (\ref{eq6-130}) and (\ref{eq6-160}), contains two terms: one
proportional to $\delta (\lambda)$ and the other to $J_1(\lambda)$. The
$\propto\delta (\lambda)$ term has no contribution to $K_{\alpha}^{(0)}$,
resulting in that both two one-sided values $H_{\alpha}^{(0)+}$ and
$H_{\alpha}^{(0)-}$ yield the same function $K_{\alpha}^{(0)}$
(for that reason, no $\pm$ sign has been assigned to this quantity).
There is an apparent danger of divergence coming from the
$O(\epsilon^{-2})$ term in Eq.\ (\ref{eq7-130}). Such a divergence
is avoided, however, as we have
\begin{equation}\label{eq7-150}
\int_{0}^{\widetilde\infty}K_{\alpha}^{(0)}(x)dx =
\frac{u_{\alpha}}{r_0^3}
\int_{0}^{\widetilde\infty}dx\,x \int_x^{\widetilde\infty}
J_1(\lambda)d\lambda=
\frac{u_{\alpha}}{2r_0^3}\int_0^{\widetilde\infty}x^2 J_1(x)dx
=0,
\end{equation}
where in the second equality we integrated by parts with respect
to $x$, and where the vanishing of the last integral is implied
by Eq.\ (\ref{eq7-5}).
Thus, the $O(\epsilon^{-2})$ term in Eq.\ (\ref{eq7-130}) vanishes,
and the $\epsilon\to 0$ limit in this equation turns out well defined.
The remaining $O(\epsilon^0)$ contribution reads
\begin{equation}\label{eq7-160}
D_{\alpha}^{(0)}=\frac{q^2 r_0}{24}\int_0^{\widetilde\infty}
[K_{\alpha}^{(0)}(x)]''dx=
-\frac{q^2r_0}{24}\int_0^{\widetilde\infty}\left\{
H_{\alpha}^{(0)\pm}(x)+\left[xH_{\alpha}^{(0)\pm}(x)\right]'
\right\}dx=-\frac{q^2r_0}{24}\int_0^{\widetilde\infty}
H_{\alpha}^{(0)\pm}(x)dx,
\end{equation}
as the surface term vanishes. Substituting for $H_{\alpha}^{(0)\pm}(x)$
and integrating using Eq.\ (\ref{eq7-5}), we finally obtain
\begin{equation}\label{eq7-170}
D_{\alpha}^{(0)}=-\frac{q^2u_{\alpha}}{24r_0^2}\int_{0}^{\widetilde\infty}
J_1(\lambda)d\lambda = -\frac{q^2u_{\alpha}}{24r_0^2}.
\end{equation}

We next turn to calculate $D_{\alpha}^{(1)}$. Writing Eq.\ (\ref{eq7-100})
for $n=1$ in the form
\begin{equation}\label{eq7-180}
D_{\alpha}^{(1)}= -q^2 r_0\lim_{\epsilon\to 0}\left[(r_0/\epsilon)
\sum_{l=0}^{\widetilde\infty}(\epsilon/r_0)\int_{L\epsilon/r_0}^
{\widetilde\infty}H_{\alpha}^{(1)}(\lambda) d\lambda\right]
\end{equation}
and applying the summation formula (\ref{EQ7-110}), we obtain
\begin{equation}\label{eq7-190}
D_{\alpha}^{(1)}= -q^2 r_0\lim_{\epsilon\to 0}\left[(r_0/\epsilon)
\int_{0}^{\widetilde\infty}dx \int_{x}^{\widetilde\infty}
H_{\alpha}^{(1)}(\lambda)d\lambda\right],
\end{equation}
with all $O(\epsilon^{2})$ terms appearing in Eq.\ (\ref{EQ7-110})
vanishing in the limit $\epsilon\to 0$.
By integrating the last expression by parts with respect to $x$
(noticing the vanishing of the surface term) one finds
\begin{eqnarray}\label{eq7-200}
D_{\alpha}^{(1)}=
-q^2 r_0\lim_{\epsilon\to 0}\left[(r_0/\epsilon)
\int_{0}^{\widetilde\infty}x H_{\alpha}^{(1)}(x) dx \right].
\end{eqnarray}
Here, the integrand contains only terms of the form $\propto x^k J_n(x)$,
with $k-n$ being positive odd integers [see Eqs.\ (\ref{eq6-140}) and
(\ref{eq6-170})]. Hence, by virtue of Eq.\ (\ref{eq7-5}), the integral
vanishes, yielding
\begin{equation}\label{eq7-210}
D_{\alpha}^{(1)}= 0.
\end{equation}

Finally, from Eq.\ (\ref{eq7-100}) with $n=2$ we obtain
\begin{eqnarray}\label{eq7-220}
D_{\alpha}^{(2)}=-q^2 r_0\lim_{\epsilon\to 0}\left[
\sum_{l=0}^{\widetilde\infty}(\epsilon/r_0)(L\epsilon/r_0)^{-1}
\int_{L\epsilon/r_0}^{\widetilde\infty} H_{\alpha}^{(2)}(\lambda)
d\lambda\right]= -q^2 r_0\int_{0}^{\widetilde\infty}(dx/x)
\int_{x}^{\widetilde\infty}H_{\alpha}^{(2)}(\lambda) d\lambda,
\end{eqnarray}
which, after integrating by parts, becomes
\begin{equation}\label{eq7-225}
D_{\alpha}^{(2)}=-q^2 r_0\left[\int_{0}^{\widetilde\infty}\ln x\,
H_{\alpha}^{(2)}(x)dx
-\lim_{x\to 0} \left(\ln x\int_{x}^{\widetilde\infty}
H_{\alpha}^{(2)}(\lambda) d\lambda\right)\right].
\end{equation}
We notice here that the second integral on the RHS is just $C_{\alpha}+O(x)$
[up to a multiplicative constant; see Eq.\ (\ref{eq6-210C})].
The above-deduced vanishing of the parameter $C_{\alpha}$ guarantees
the definiteness of the $x\to 0$ limit in Eq.\ (\ref{eq7-225}), and
makes the second term on its RHS vanish.
Note the way the vanishing of the parameter $C_{\alpha}$ appears as
a necessary self-consistency condition in our scheme: had we got
$C_{\alpha}\ne 0$, the parameter $D_{\alpha}$ would have been indefinite,
and the whole regularization scheme would have been rendered meaningless.

As the $x\to 0$ limit in Eq.\ (\ref{eq7-225}) vanishes, we are left with
\begin{equation}\label{eq7-230}
D_{\alpha}^{(2)}=-q^2 r_0\int_{0}^{\widetilde\infty}\ln x\,
H_{\alpha}^{(2)}(x) dx.
\end{equation}
With the explicit form of $H_{\alpha}^{(2)}$, given in Eqs.\
(\ref{eq6-150}) and (\ref{eq6-180}), the integrand in the last expression
is found to consist of various terms of the form
$\propto x^k \ln x\, J_n(x)$, with $k-n$ being positive odd integers.
The integrals $\int_{0}^{\widetilde\infty}$ of such terms can be read
from the formula (\ref{eq7-6}), which we derive in Appendix \ref{AppB}.
Using this formula we obtain expressions for $D_r^{(2)}$ and $D_t^{(2)}$,
which, after
substituting for $\dot{r}_*$, $\ddot{r}_*$, $\overdots{r}_*$, $\ddot{s}$,
$\overdots{s}$, $f_1$, $f_2$, and $f_3$, read
\begin{eqnarray}\label{eq7-250}
D_r^{(2)}=\frac{1}{3}q^2\left(f^{-1}\overdots{r}+\dot{r}\ddot{v}\ddot{u}
\right) +\frac{q^2\dot{r}}{24f^2r_0^2}\left[f(4f-3)+8r_0f'(f-r_0\ddot{r})
+2r_0^2(3ff''-{f'}^2)\right],
\end{eqnarray}
and
\begin{eqnarray}\label{eq7-255}
D_t^{(2)}&=&-\frac{1}{3}q^2\left(f\overdots{r}+f^2\dot{t}\ddot{v}\ddot{u}
\right)-\frac{1}{2}q^2f'\dot{r}\ddot{t}\nonumber\\
&&+\frac{q^2\dot{t}}{24r_0^2}\left[3f-4f^2+4f^{-1}(f')^2r_0^2\dot{r}^2
-8f''r_0^2\dot{r}^2-4f'r_0^2\ddot{r}-2ff''r_0^2-8ff'r_0\right].
\end{eqnarray}

We are now in position to write an expression for the ``overall''
parameter $D_{\alpha}$.
We have $D_{\alpha}=D_{\alpha}^{(0)}+D_{\alpha}^{(2)}$, yielding
\begin{eqnarray}\label{eq7-290}
D_r=\frac{1}{3}q^2\left(f^{-1}\overdots{r}+
\dot{r}\ddot{v}\ddot{u}\right)
+\frac{q^2\dot{r}}{12f^2r_0^2}\left[2f(f-1)+4r_0f'(f-r_0\ddot{r})
+r_0^2(3ff''-{f'}^2)\right],
\end{eqnarray}
and
\begin{eqnarray}\label{eq7-300}
D_t&=&-\frac{1}{3}q^2\left(f\overdots{r}+f^2\dot{t}\ddot{v}\ddot{u}\right)
-\frac{1}{2}q^2f'\dot{r}\ddot{t}\nonumber\\
&&+\frac{q^2\dot{t}}{12r_0^2}\left[2f(1-f)+2f{-1}(f')^2r_0^2\dot{r}^2
-4f''r_0^2\dot{r}^2-2f'r_0^2\ddot{r}-ff''r_0^2-4ff'r_0\right].
\end{eqnarray}

In the case of radial motion, the four-acceleration's
components admit the explicit form $a^r=\ddot{r}+\frac{1}{2}f'$ and
$a^t=\ddot{t}+f'\dot{r}_*\dot{t}$. Recalling also that in the spacetime
class considered here the Ricci scalar reads $R=-[f''+4f'/r+2(f-1)/r^2]$,
one can show that the above two expressions for $D_r$ and $D_t$
can be put into the simple vectorial form
\begin{equation}\label{eq7-310}
D_{\alpha}=\frac{1}{3}q^2 \left(\dot{a}_{\alpha}-a^2 u_{\alpha}\right)
-\frac{1}{12}q^2Ru_{\alpha}.
\end{equation}
Comparing now this result with Eqs.\ (\ref{eq2-100}) and (\ref{eq2-110}),
and recalling that in the case considered here (that of radial motion
on static spherically-symmetric background) the first two
terms in the expression for $F_{\alpha}^{\rm (Ricci)}$ [Eq.\
(\ref{eq2-110})] cancel out, we arrive at the remarkable conclusion that
$D_{\alpha}$ is exactly the ``standard'' local part of the self-force:
\begin{equation}\label{eq7-320}
D_{\alpha}=F_{\alpha}^{\rm (ALD)}+F_{\alpha}^{\rm (Ricci)}.
\end{equation}

\section{Summary and concluding remarks}\label{secVIII}

The total self force acting on the radially moving scalar particle
is obtained by substituting $F^{\rm (tail)}_{\alpha}$ from Eq.\
(\ref{eq6-230}) in Eq.\ (\ref{eq2-90}). By virtue of Eq.\ (\ref{eq7-320}),
the contribution of $D_{\alpha}$ to the tail term is then found to
exactly {\it cancel out} the local term in the expression for the total
self force. This, in addition to the vanishing of the parameter
$C_{\alpha}$, leads to the simple result
\begin{equation}\label{eq8-10}
F^{\rm (total)}_{\alpha}=\sum_{l=0}^{\infty}\left(F^{l\pm}_{\alpha}
-A_{\alpha}^{\pm}L-B_{\alpha}\right)
\end{equation}
(where, we recall, $L=l+1/2$).
An even simpler form is obtained when calculating $F^{\rm (total)}_{\alpha}$
using the {\em averaged} value of the modes $F_{\alpha}^l$, obtained
by averaging over their two one-sided values. Then, by virtue of
Eq.\ (\ref{eq7-20}), we find
\begin{equation}\label{eq8-20}
F^{\rm (total)}_{\alpha}=\sum_{l=0}^{\infty}\left(\bar F^l_{\alpha}
-B_{\alpha}\right),
\end{equation}
where $\bar F^l_{\alpha}\equiv\frac{1}{2}(F^{l+}_{\alpha}+F^{l-}_{\alpha})$.
Recall that the parameter $B_{\alpha}$, given in Eq.\ (\ref{eq7-70}),
is just the asymptotic value of the
averaged $l$-mode $\bar F^l_{\alpha}$ at the limit $l\to\infty$;
namely, the total self force is obtained by simply subtracting from
each (two one-sided averaged) mode its large $l$ asymptotic value,
and then summing over all modes.\footnote{
The simplicity of our main result, Eq.\ (\ref{eq8-20}), may lead
one to wonder whether there could be simple arguments leading directly to
this result. Such arguments might perhaps rely on general properties
of the Hadamard expansion. This should make an interesting subject
for further investigation.}

To summarize, in this paper we have developed a method for calculating the
self force on a scalar particle in curved spacetime, through regularization
of the multipole mode sum. The basic difficulty in applying the mode
decomposition approach---the apparent divergence of the sum over
modes---has been taken care of by the introduction of an appropriate
regularization scheme,
providing a practical prescription for calculating the self force.
It should be emphasized that the proposed method does not involve
any weak-field or slow-motion approximations, and thus allows effective
calculations of the self force even for strong field orbits.

The basic expression for the tail part of the self force is given in
Eq.\ (\ref{eq6-230}). This expression was developed here for
radial motion; however, the same general form applies for any
trajectory \cite{unpublished}, with the details of the
orbit encoded only in the values of the regularization parameters
(as well as, of course, in the form of the ``bare'' modes
$F_{\alpha}^{l\pm}$).
To apply this general expression for a given trajectory requires
knowledge of
(i) the modes $F^l_{\alpha}$, to be derived by supplementary
(basically straightforward) numerical analysis, as done in Refs.\
\cite{Burko1,Burko2,BB}; and (ii) four regularization parameters
for each spacetime component of the force.
In this paper we have worked out the entire calculation
of the regularization parameters for the case of radial motion.
For any other trajectory, the derivation of these parameters can be
carried out along the same lines, based on the explicit form of the
Green's function's $l$-mode given in Eq.\ (\ref{eq3-150}),
with Eqs.\ (\ref{eq5-150}), (\ref{eq5-170}), and (\ref{eq5-190}).
To that end,
one first obtains an expression for the gradient of the Green's
function's $l$-mode, as in Eq.\ (\ref{eq5-230}) [supplemented with
Eqs.\ (\ref{eq5-250}) and (\ref{eq5-260})].
One next expands this gradient in powers of proper time $\tau$
along the worldline about the force's evaluation point, and
re-expresses the resulting expansion as an expansion in powers of
$1/L$, by holding $\tau L$ fixed---as in Eq.\ (\ref{eq6-120})
[supplemented with Eqs.\ (\ref{eq6-130})--(\ref{eq6-180})].
One finally uses the values of the above expansion coefficients
[denoted in this paper by $H_{\alpha}^{(n)}(\tau L)$] to construct
the regularization parameters through Eqs.\ (\ref{eq6-210}) and
(\ref{eq6-220}).

For a scalar particle moving radially in a spacetime of the class
considered in this paper, we found the total self force to be given
by Eq.\ (\ref{eq8-20}). This constitutes our main result for the
radial motion case, together with the explicit values of the
regularization parameters given in Eqs.\ (\ref{eq7-20}),
(\ref{eq7-70}), (\ref{eq7-80}), and (\ref{eq7-310}).
We have found that, in the radial motion case, the parameters
$\bar A_{\alpha}$ and $C_{\alpha}$ both vanish [the one-sided values
of $A_{\alpha}$ do not vanish; they are given in Eq.\ (\ref{eq7-10})],
and the parameter $D_{\alpha}$ is just the standard local part of the self
force. The vanishing of $C_{\alpha}$, shown here explicitly, appeared
as a necessary condition for the definiteness of the whole scheme
(had $C_{\alpha}\neq 0$, the parameter $D_{\alpha}$ would have diverge;
see the discussion in Sec.\ \ref{secVII}). This point serves to
demonstrate the self-consistency of the regularization scheme.

A question arises, whether the above results (the vanishing of
$\bar A_{\alpha}$ and  $C_{\alpha}$ and the special value of $D_{\alpha}$)
represent generic features of the regularization scheme, or rather are
special to radial trajectories. Preliminary investigation \cite{unpublished}
suggests that, indeed, $\bar A_{\alpha}$ and  $C_{\alpha}$ vanish for all
trajectories, at least in the Schwarzschild case.
As to the parameter $D_{\alpha}$, this was shown so far to obey Eq.\
(\ref{eq7-320}) at least in one more important example, that
of a circular orbit around a Schwarzschild black hole
\cite{unpublished,Rapid}. It might be conjectured (and be subject
to further investigation) that Eq.\ (\ref{eq7-320}) holds for any
trajectory in any static spherically-symmetric background. In that case,
the simple Eq.\ (\ref{eq8-20}) for the total self force would be valid
for all such trajectories.

Under the above conjecture, we find that regularization of the
total self force requires knowledge of just one parameter,
$B_{\alpha}$, representing the asymptotic value of the modes
$\bar F_{\alpha}^l$ as $l\to\infty$.
The value of $B_{\alpha}$ for any specific radial trajectory on any
given spherically-symmetric spacetime, can be read from
Eq.\ (\ref{eq7-70}) (valid regardless of the above conjecture).
For example, in the special case of a {\em static} particle we
find $B_{t}^{\rm (static)}=0$ and
\begin{equation}\label{eq8-30}
B_{r}^{\rm (static)}=
-\frac{q^2}{2r_0^2}\left(1+\frac{r_0f'}{2f}\right).
\end{equation}
For radial {\em geodesic} motion we find
\begin{equation}\label{B}
B_t^{(\rm geodesic)}=-\frac{q^2}{2r_0^2}\,\dot{r}\,E,
\quad\quad
B_r^{(\rm geodesic)}=-\frac{q^2}{2r_0^2}\left(2-E^2/f\right),
\end{equation}
where $E\equiv -u_t$ is the energy parameter (which is a constant of
motion in the absence of self force effect).
For the value of $B_{r}$ in the case of uniform {\em circular} motion in
Schwarzschild spacetime (the derivation of which will be presented
elsewhere \cite{unpublished}) we refer the reader to Eq.\ (34) of
Ref.\ \cite{Rapid}.

The applicability of the regularization prescription described here was
demonstrated recently in actual calculations of the self force for various
scenarios. Burko first studied the cases of static \cite{Burko1}
and circular \cite{Burko2} orbits in the Schwarzschild spacetime.
For these stationary scenarios, the modes $F_{\alpha}^l$ were obtained by
summing over the Fourier-multipole modes $F_{\alpha}^{lm\omega}$, first
derived by solving the appropriate ordinary field equations in the
frequency domain.
Later, Barack and Burko \cite{BB} analyzed the case of
radial motion in Schwarzschild. In this case, which is no longer
stationary, numerical evolution of the appropriate partial DE in
the time domain was applied to directly infer the modes $F_{\alpha}^l$.
In each of these studies, the overall force acting on the scalar
particle was finally deduced by summing over all modes, using the
above regularization scheme.
In each of the cases analyzed, the vanishing of $C_{\alpha}$ was
demonstrated, and the analytically-derived expressions
for $A_{\alpha}$ and $B_{\alpha}$ were verified.
Aside from demonstrating the applicability of the regularization scheme
and providing verification for the values of the regularization
parameters, the above studies yielded valuable physical results, as
mentioned in the Introduction.

Of course, the analysis of the scalar self force merely serves as
a toy model for more realistic cases.
Generalization of the regularization scheme to the {\em electromagnetic}
self force seems possible, based on the existing formalism \cite{DB,QW1}.
Such a generalization is necessary, for example,
for resolving the interesting question re-raised recently by Hubeny
\cite{Hubeny}, whether a nearly extreme electrically-charged black
hole might be overcharged (and its event horizon by destroyed) by
throwing in a charged particle: as pointed out by Hubeny, knowledge
of the exact radiation reaction effect is crucial for obtaining a
definite answer. More difficult to accomplish would be the important
generalization of the scheme to the gravitational self force acting
on a mass particle.

Finally, it should be mentioned that a closely related approach
was recently applied by Lousto \cite{Lousto} for studying the
gravitational self force on a mass particle in Schwarzschild
spacetime. This approach is also based on the multipole expansion,
yet it employs a different regularization method for the mode sum
(it is argued that the correct self force can be deduced
by applying the zeta-function regularization technique).
For the geodesic motion case studied by Lousto, this approach
leads to an expression analogous to Eq.\ (\ref{eq8-20}).

\section*{acknowledgements}

I wish to thank Amos Ori for suggesting the basic idea for the
regularization scheme, and for his assistance in developing it.
I would also like to thank Lior Burko for discussions and for
reading the manuscript.

\appendix

\section{Multipole expansion of the Green's function
using ``tilde-summation''}\label{AppA}

DeWitt and Brehme wrote a general expression for the scalar Green's
function in curved spacetime [see Eq.\ (2.21) of Ref.\ \cite{DB}],
of the form
\begin{equation}\label{eqA-10}
G(x^{\mu};{x'}^{\mu})= a(x^{\mu};{x'}^{\mu})\,\delta(\sigma)+
b(x^{\mu};{x'}^{\mu})\,\Theta(\sigma).
\end{equation}
Here, $\sigma$ is plus or minus half the squared geodesic distance
between the source point $x^{\mu}$ and the evaluation point ${x'}^{\mu}$,
according to whether the geodesic connecting the points (along which
the invariant distance is measured) is timelike ($\sigma>0$) or else
($\sigma\leq 0$);
$a$ is a certain function having a well defined value at $\sigma=0$;
and $b$ is a function which may be written as a Taylor expansion in
$\sigma$ about $\sigma=0$. [This expansion was shown by Hadamard
(see pp.\ 96--98 in Ref.\ \cite{Hadamard}) to converge uniformly at
least inside the region where $\sigma$ is single valued.]
While the first term in Eq.\ (\ref{eqA-10}) is associated with the
familiar delta function exhibited already in flat spacetime, the second
term represents a curvature-induced tail, which ``fills'' the light
cone (defined by $\sigma=0$). Note that the Green's function is
strongly irregular along the light cone of the source point
${x'}^{\mu}$.

Now, the question to consider is whether the above Green's function
may be expanded in terms of the standard spherical harmonic functions
$Y^{lm}(\theta,\varphi)$ on a sphere of constant $r,t$.
Standard theorem (see, e.g., Ref.\ \cite{Hilbert}, p.\ 513)
states that a {\em sufficient} condition for (absolute and uniform)
convergence of the spherical harmonic expansion is the expanded
function being $C^2$ on the sphere. This condition is not satisfied
here, as the Green's function diverges along the curve generated
by the intersection of the future light cone of ${x'}^{\mu}$ and
the sphere of constant $r,t$. Therefore, it is not guaranteed,
in advance, that such an expansion could be naively applied.
Indeed, already in flat spacetime the attempt to apply the multipole
expansion to the Green's function turns out to yield a divergent
sum---see Eq.\ (\ref{eq3-4}) and  the discussion proceeding it in
Sec.\ \ref{secIII}.

Let us introduce the ``modified'' Green's function
(which is not a``Green's function'' for the scalar field anymore),
defined by
\begin{equation}\label{eqA-20}
G_{\rm mod}=G-\delta G,
\end{equation}
where
\begin{equation}\label{eqA-30}
\delta G\equiv a_0\,\delta(\sigma) - b_0\,\Theta(-\sigma),
\end{equation}
with $a_0\equiv a(\sigma=0)$ and $b_0\equiv b(\sigma=0)$.
The function $G_{\rm mod}$ has two essential features: (i) its support
inside the light cone is identical to that of the Green's function $G$
(as $\delta G$ has no support there), and (ii) it is continuous throughout
any sphere of constant $r,t$.
The first feature implies that we can use $G_{\rm mod}$ instead of
$G$ in calculating the self force: We may re-write Eq.\ (\ref{eq2-130})
as
\begin{equation}\label{eqA-40}
F^{(\epsilon)}_{\alpha}\equiv q^2\int_{-\infty}^{-\epsilon}
\left\{G_{{\rm mod}}\left[x^{\mu}_0;x_p^{\mu}(\tau)\right]\right\}_
{,\alpha} d\tau,
\end{equation}
as the $\delta G$ term in $G_{\rm mod}$ contributes nothing to
the integral along the particle's worldline.

The second of the above features of $G_{\rm mod}$, its continuity,
may imply that the multipole expansion could now be applied to it.
Strictly speaking, $G_{\rm mod}$ does not satisfy the above sufficient
condition for absolute and uniform convergence of the mode sum
(i.e., being $C^2$ on the sphere); yet, we shall {\em assume} here
that $G_{\rm mod}$, being continuous, is already regular enough to
admit a convergent mode sum. The results of our analysis turn out
to be consistent with this assumption, as the mode-sum of $G$
considered (based on the function $G_{\rm mod}$ through the use
of ``tilde summation''---see below) is found to be (absolutely)
convergent. The validity of this assumption can also be demonstrated
in the flat space case: the $\propto b_0$ tail term vanishes in this
case, and by expanding the $\delta(\sigma)$ term in Eq.\ (\ref{eqA-30})
in spherical harmonics one can easily verify that the $l$-mode
of $\delta G$ is exactly the $l$-mode of $G$, given in Eq.\ (\ref{eq3-4}).
The $l$-mode of $G_{\rm mod}$ then vanishes, and the mode sum converges.
Although trivial, this flat-space example may serve to demonstrate
how the subtraction of $\delta G$ from $G$ already removes the
divergent piece from the $l$-mode, making the mode sum
converge.\footnote{
One may similarly construct a more sophisticated
function $\delta G$, designed to yield a $C^2$ modified
function $G_{\rm mod}$ [by canceling also the $O(\sigma)$ and
$O(\sigma^2)$ terms in the Taylor expansion of $b$] being sufficiently
regular to {\em assure} uniform and absolute convergence of the multipole
expansion, by standard mathematical theorem. Such an improved
construction will not be examined here.
}
We thus expand $G_{\rm mod}$ as
\begin{equation}\label{eqA-50}
G_{\rm mod}=\sum_{l=0}^{\infty} G_{\rm mod}^l=
\sum_{l=0}^{\infty} \left(G^l-\delta G^l\right),
\end{equation}
where $G_{\rm mod}^l$ and $\delta G^l$ are the spherical harmonic
modes of $G_{\rm mod}$ and $\delta G$, respectively
(obtained by summing over all azimuthal numbers $m$).

To proceed, let us now define the new operation
$\widetilde\lim_{l\to\infty}$ (``tilde-limit'') as follows:
Consider a series of numbers $A_l$ (with $l=0,1,\ldots,\infty$).
Let $B_l^{(j)}$ be any expression of the form
\begin{equation}\label{eqA-60}
B_l^{(j)}=a_j l^{b_j}\cos(\alpha_j l+\beta_j),
\end{equation}
where $a_j$, $b_j$, $\alpha_j$, and $\beta_j$ are some $l$-independent
real numbers, with $\alpha_j\neq 0$ for all $j$.
If there exists a finite number $k$ of expressions $B_l^{(j)}$ of
this form (with $j=1,2,\ldots,k$), such that subtracting their sum from
the original series $A_l$ would yield a well-defined finite limit as
$l\to\infty$, then we define the ``tilde-limit''
$\widetilde\lim_{l\to\infty}A_l$ as in Eq.\ (\ref{eq3-7}).

One may easily be convinced that the tilde-limit is single-valued
(when existing). For, suppose that for a given series $A_l$ there
were two different sets of quantities $B_l^{(j)}$, one
(denoted by $\bar B_l^{(\bar j)}$) yielding
$\widetilde\lim_{l\to\infty}A_l=\bar c$, and the other (denoted by
$\hat B_l^{(\hat j)}$) yielding $\widetilde\lim_{l\to\infty}A_l
=\hat c\neq \bar c$. Then, for the difference between the two
limits one would have found
$\lim_{l\to\infty}\left[\sum_{\hat j} \hat B_l^{(\hat j)}-
\sum_{\bar j} \bar B_l^{(\bar j)}\right]=\hat c-\bar c\neq 0$.
This, however, is impossible, as the (standard) limit $l\to\infty$
of any quantity of the type $B_l^{(j)}$ is either diverging or zero,
and so is the limit of any finite sum of such quantities.
Hence, we must have $\hat c=\bar c$, and the tilde-limit is single
valued. In particular, we find that if there exists a finite standard
limit $\lim_{l\to\infty} A_l$, then
$\lim_{l\to\infty} A_l=\widetilde\lim_{l\to\infty}A_l$.

We can now also define the ``tilde-sum'' of a series $A_l$,
as in Eq.\ (\ref{eq3-8}).
Again, if the ``tilde-sum'' of a series exists, then it is {\em unique}.
Also, if the standard sum $\sum_{l\to\infty}A_l$ converges, then
we may replace it with a ``tilde-sum'' operation. In particular, we may
replace the convergent standard sum of Eq.\ (\ref{eqA-50}) with a
tilde-summation:
\begin{equation}\label{eqA-90}
G_{\rm mod}=\sum_{l=0}^{\widetilde\infty}\left(G^l-\delta G^l\right).
\end{equation}
Below we show that
\begin{equation}\label{eqa-100}
\sum_{l=0}^{\widetilde\infty} \delta G^l=0
\end{equation}
(for any evaluation point $x_0^{\mu}$ lying inside the future light
cone of the source point $x_{p}^{\mu}$).
As a consequence, the tilde-sum of $G^l$ is found to be finite
and equal to $G_{\rm mod}$---as indicated in Eq.\ (\ref{eq3-10})
of Sec.\ \ref{secIII}.

Combining Eqs.\ (\ref{eqA-40}) and (\ref{eq3-10}) we conclude
that the self-force can be calculated by analyzing the modes $G^l$
of the {\em original} Green's function, provided that in order
to sum over all modes one applies the {\em tilde}-summation
instead of the standard summation.
The validity of this statement crucially depends on the vanishing
of the tilde-sum over $\delta G^l$ [Eq.\ (\ref{eqa-100})],
which we now prove.

\subsection*{Proof of Eq.\ (\ref{eqa-100})}

To calculate the $l$-modes of $\delta G$, it is convenient to
use a spherical coordinate system in which the source point
$x_{\rm p}^{\mu}$ lies on the polar axis (i.e., $\theta_p=0$).
In this coordinate system, contributions to $\delta G$ would
come only from the $m=0$ modes:
\begin{eqnarray}\label{eqA-110}
\delta G^l &=&\sum_{m=-l}^l Y_{lm}(\theta,\varphi)
\int_0^{2\pi}d\varphi'\int_{-1}^1d(\cos\theta')\,
\delta G(\sigma')\,
Y_{lm}^*(\theta',\varphi')=\nonumber\\&&
L\int_{-1}^1d(\cos\theta')
\left[a_0\,\delta(\sigma')-b_0\,\Theta(-\sigma')\right]\,
P_l(\cos\theta) P_l(\cos\theta'),
\end{eqnarray}
where the integration is carried out over a sphere spanned by
$\theta',\varphi'$ (containing the evaluation point $x^{\mu}$),
$L\equiv (l+1/2)$,
and $\sigma'$ is half the squared geodesic distance between the
source point and the integration point $\theta',\varphi'$.
Now, the future light cone of the source point (along which $\sigma'=0$)
intersects the integration sphere along a circle
$\theta'={\rm const}\equiv\theta_0$. The $\delta(\sigma')$ term in Eq.\
(\ref{eqA-110}) contributes to the integration only along this
circle, while the $\Theta(-\sigma)$ term contributes only across the
part of the sphere outside it. We thus find
\begin{equation}\label{eqA-115}
\delta G^l=L P_l(\cos\theta)
\left[\hat a_0 P_l(\cos\theta_0)-b_0
\int_{-1}^{\cos\theta_0}P_l(\cos\theta')\,d(\cos\theta')\right]
\equiv \delta G^l_{(\delta)}+\delta G^l_{(\Theta)},
\end{equation}
where $\hat a_0\equiv a_0/
\left|d\sigma'/d(\cos\theta')\right|_{\theta'=\theta_0}$
is a constant, and where the symbols $\delta G^l_{\delta}$ and
$\delta G^l_{\Theta}$ represent the two terms proportional to
$\hat a_0$ and $b_0$, respectively.

To calculate the tilde-sum over $\delta G^l$ we make use of the
finite-sum identity [referred to as ``Christoffel's first summation
formula''---see, e.g., Eq.\ 3.8(20) of Ref.\ \cite{Erdelyi}]
\begin{equation}\label{eqA-120}
(x-y)\sum_{l'=0}^{l} LP_{l'}(x)P_{l'}(y)=\frac{1}{2}\,
(l+1)\left[P_{l+1}(x)P_{l}(y)-P_{l+1}(y)P_{l}(x)\right],
\end{equation}
valid for $|x|\leq 1$ and $|y|\leq 1$.
Applying this formula to Eq.\ (\ref{eqA-115}), we obtain
\begin{eqnarray}\label{eqA-130}
\sum_{l=0}^{\widetilde\infty}\delta G^l&=&
\widetilde\lim_{l\to\infty}\sum_{l'=0}^{l}
\left(\delta G^{l'}_{(\delta)}+\delta G^{l'}_{(\Theta)}\right)=
\nonumber\\&&
\frac{1}{2}\,\hat a_0\,\widetilde\lim_{l\to\infty}
\left[(l+1) \frac{P_{l+1}(x)P_{l}(x_0)-
P_{l+1}(x_0)P_{l}(x)}{x-x_0}\right]
\nonumber\\&&
-\frac{1}{2}\, b_0\,\widetilde\lim_{l\to\infty}
\left[(l+1)\int_{-1}^{x_0}\frac{P_{l+1}(x)P_{l}(x')-
P_{l+1}(x')P_{l}(x)}{x-x'}\,dx'\right],
\end{eqnarray}
where $x$ and $x_0$ stand for $\cos\theta$ and $\cos\theta_0$.
(Recall that for any evaluation point $x^{\mu}$ lying inside the
future light cone of the source point $x_{p}^{\mu}$, we have
$\cos\theta>\cos\theta_0\geq\cos\theta'$; hence, the denominators
appearing in the last expression are strictly positive.)

Let us consider first the tilde-sum of $\delta G^l_{(\delta)}$.
For values of $\theta$ satisfying $\epsilon\leq \theta\leq \pi-\epsilon$
(where $\epsilon>0$) we have the large-$l$ asymptotic form
\begin{equation}\label{eqA-150}
P_l(\cos\theta)\propto
\sqrt{2}\,(l\,\pi\sin\theta)^{-1/2}\,\cos(L\theta+\pi/4)+O(l^{-3/2})
\end{equation}
[see, e.g., Eq.\ 3.9(2) of Ref.\ \cite{Erdelyi}].
Using this asymptotic form with Eq.\ (\ref{eqA-130}), we may easily
write $\sum_{l'=0}^{l}\delta G^{l'}_{(\delta)}$ as a sum
of a few terms of the form
$
a_j\cos(\alpha_j l + \beta_j)+O(l^{-1})
$
(in the case $0<\theta<\pi$),
or of the form
$
a_jl^{1/2}\cos(\alpha_j l + \beta_j)+O(l^{-1/2})
$
(in the case $\theta=0$ or $\pi$),
where $a_j$, $\alpha_j\neq 0$, and $\beta_j$ are certain functions
of $\theta$ and $\theta_0$ (independent of $l$). Such terms all
vanish at the tilde-limit $l\to\infty$. Thus, we clearly have
\begin{equation}\label{eqA-190}
\sum_{l=0}^{\widetilde\infty}\delta G^l_{(\delta)}=0.
\end{equation}

We next turn to calculate the tilde-sum of $\delta G^l_{(\Theta)}$.
Integrating by parts in Eq.\ (\ref{eqA-130}), and using
\begin{equation}\label{eqA-200}
\int P_l(x)\,dx=\frac{P_{l+1}(x)-P_{l-1}(x)}{2l+1}
\end{equation}
[see Eq.\ 7.111 of Ref.\ \cite{GradRyz}, together with Eq.\ 8.733-4
therein], we obtain
\begin{eqnarray}\label{eqA-210}
\sum_{l'=0}^{l}\delta G^{l'}_{(\Theta)}=
-\frac{1}{2}\,b_0(l+1)  &\left\{
\frac{P_{l+1}(x)}{2l+1}\left[\frac{P_{l+1}(x')-P_{l-1}(x')}
{x-x'}\right|_{-1}^{x_0}\left.-\int_{-1}^{x_0} \frac{P_{l+1}(x')-
P_{l-1}(x')}{(x-x')^2}\,dx' \right]\right.
\nonumber\\
&-\left.\frac{P_{l}(x)}{2l+3}\left[\frac{P_{l+2}(x')-P_{l}(x')}
{x-x'}\right|_{-1}^{x_0}\left.-\int_{-1}^{x_0} \frac{P_{l+2}(x')-
P_{l}(x')}{(x-x')^2}\,dx' \right]
\right\}.
\end{eqnarray}
The surface terms here vanish at the lower boundary,
as $P_{ l+1}(-1)=P_{l-1}(-1)$ and
$P_{l+2}(-1)=P_{l}(-1)$ [recalling $P_l(-1)=(-1)^l$].
The contribution from the upper boundary dies off at large $l$
[by virtue of Eq.\ (\ref{eqA-150})] as $\propto l^{-1}$ (times
oscillations) for $\theta\ne 0$, or as $\propto l^{-1/2}$ (times
oscillations) for $\theta=0$. In both cases, this contribution thus
vanishes at the standard limit $l\to\infty$, and hence also at the
tilde-limit. Consider next the integral terms:
The difference between two Legendre functions appearing in these terms
may be globally bounded (in absolute value) as
\begin{equation}\label{eqA-215}
\left|P_{l+1}(\cos\theta)-P_{l-1}(\cos\theta)\right|\leq
C_0[\pi (l-1)]^{-1/2},
\end{equation}
where $C_0$ is a number independent of $l$ and $\theta$
[see Eq.\ 8.838 of Ref.\ \cite{GradRyz}].
For any $\theta$ and $\theta_0$, the integrals of Eq.\ (\ref{eqA-210})
are thus each bounded (in absolute value) by
$C_1(\theta,\theta_0)\times l^{-1/2}$, and we find
\begin{equation}\label{eqA-220}
\left|\sum_{l'=0}^{l}\delta G^{l'}_{(\Theta)}\right|\leq
C_2(\theta,\theta_0)\,l^{-1/2}\left[\left|P_{l+1}(\cos\theta)\right|
+\left|P_{l}(\cos\theta)\right|\right] \to 0
\end{equation}
as $l\to\infty$
(the above coefficients $C_1$ and $C_2$ are $l$-independent).
Thus, the standard infinite sum over $\delta G^l_{(\Theta)}$ vanishes,
and hence also the tilde-sum:
\begin{equation}\label{eqA-230}
\sum_{l=0}^{\widetilde\infty}\delta G^l_{(\Theta)}=0.
\end{equation}

With Eqs.\ (\ref{eqA-190}) and (\ref{eqA-230}), Eq.\ (\ref{eqa-100})
is verified.

\section{Derivation of integrals}          \label{AppB}

In this appendix we obtain the tilde-integrals given in Eqs.\
(\ref{eq7-5}) and (\ref{eq7-6}), which are needed in the calculation
of the regularization parameters. We start with Eq.\ (\ref{eq7-5}).
Let $I^k_n(\lambda)$ denote the primitive function of
$\lambda^kJ_n(\lambda)$, where $\lambda$ is a real variable, $J_n$
is the Bessel function of the first kind, of order $n$, and
$k,n\in\mathbb{N}$:
\begin{equation}\label{eqB-10}
I^k_n(\lambda)\equiv \int \lambda^kJ_n(\lambda)d\lambda.
\end{equation}
Let also $\widetilde{I}^k_n$ stand for the definite integral
\begin{equation}\label{eqB-20}
\widetilde{I}^k_n\equiv \widetilde\lim_{\lambda\to\infty}
\int_0^\lambda (\lambda')^kJ_n(\lambda')d\lambda'\equiv
\int_0^{\widetilde\infty}\lambda^kJ_n(\lambda)d\lambda.
\end{equation}

Consider first the case $k=0$. The standard integral
$\int_0^{\infty}J_n(\lambda)d\lambda=1$ is well defined and finite.
Thus, in this case, the tilde-integration in Eq.\ (\ref{eqB-20}) may
be replaced with a standard integration, yielding
\begin{equation}\label{eqB-30}
\widetilde{I}^{k=0}_n=1, \quad \forall n\geq 0.
\end{equation}

Consider next the case $k>0$.
Writing in Eq.\ (\ref{eqB-10}) $\lambda^k J_n(\lambda)=\lambda^{k-n-1}
\left[\lambda^{n+1} J_n(\lambda)\right] =
\lambda^{k-n-1}\left[\lambda^{n+1}J_{n+1}(\lambda)\right]'$
[where use is made of Eq.\ (\ref{eq5-200}) and a prime denotes
$d/d\lambda$], and integrating by parts, we arrive at the
recursive formula
\begin{equation}\label{eqB-40}
I^k_n(\lambda)=\lambda^kJ_{n+1}(\lambda)-(k-n-1)I^{k-1}_{n+1}(\lambda).
\end{equation}
If $0<k\leq n$, then by $k$ successive applications of this recursive
formula we obtain
\begin{equation}\label{eqB-60}
I^k_n(\lambda)=\sum_{j=0}^{k-1}\left[\frac{(n-k-1+2j)!!}{(n-k-1)!!}
\lambda^{k-j} J_{n+1+j}(\lambda)\right] +\frac{(n+k-1)!!}{(n-k-1)!!}
I^{0}_{n+k}(\lambda),
\end{equation}
Now, the Bessel functions $J_n(\lambda)$ admit the asymptotic form
\begin{equation}\label{eqB-200}
J_n(\lambda\to\infty)\sim (2/\pi \lambda)^{1/2}
\cos\left(\lambda-n\pi/2-\pi/4\right)
\end{equation}
(see, e.g., Eq.\ 8.451-1 in \cite{GradRyz}). Therefore, each of the $k$
terms in the sum over $j$ in Eq.\ (\ref{eqB-60}) diverges at large
$\lambda$ as some positive (half-integer) power of $\lambda$ times
oscillations with respect to
$\lambda$. Clearly, all such terms are eliminated when the tilde-limit
$\lambda\to\infty$ is taken. Also, all of these terms vanish at $\lambda=0$.
We are thus left with
\begin{equation}\label{eqB-70}
\widetilde{I}^k_n=\frac{(n+k-1)!!}{(n-k-1)!!}\,\widetilde{I}^{0}_{n+k}=
\frac{(n+k-1)!!}{(n-k-1)!!},
\quad\text{for $0<k\leq n$},
\end{equation}
where the last equality is due to Eq.\ (\ref{eqB-30}).

If $k>n$ and the difference $k-n$ is {\em even}, then by $p\equiv(k-n)/2$
applications of the recursive formula (\ref{eqB-40}) we obtain
\begin{equation}\label{eqB-80}
I^k_n(\lambda)=\sum_{j=0}^{p-1}(-1)^j\frac{(k-n-1)!!}{(k-n-1-2j)!!}
\lambda^{k-j}J_{n+1+j}(\lambda) +(-1)^p(k-n-1)!!\,I^{(k+n)/2}_{(k+n)/2}
(\lambda).
\end{equation}
Again, all terms in the sum over $j$ vanish at the tilde-limit
$\lambda\to\infty$ and at $\lambda=0$, leading to
\begin{eqnarray}\label{eqB-90}
\widetilde{I}^k_n&=&(-1)^p(k-n-1)!!\,I^{(k+n)/2}_{(k+n)/2}(\lambda)
=\nonumber\\
&&(-1)^{(k-n)/2}(k-n-1)!!(k+n-1)!!, \quad\text{for even $k-n>0$},
\end{eqnarray}
where the last equality is due to Eq.\ (\ref{eqB-70}).

The situation is different in case $k>n$ and the difference $k-n$
is {\em odd}. Then, following $q=(k-n-1)/2$ applications of the recursive
formula (\ref{eqB-40}) one obtains
\begin{eqnarray}\label{eqB-100}
I^k_n(\lambda)&=&
\sum_{j=0}^{q-1}(-1)^j\frac{(k-n-1)!!}{(k-n-1-2j)!!}
\lambda^{k-j} J_{n+1+j}(\lambda)+ (-1)^q (k-n-1)!!\,
I^{(k+n+1)/2}_{(k+n-1)/2}= \nonumber\\
&&\sum_{j=0}^{q}(-1)^j\frac{(k-n-1)!!}{(k-n-1-2j)!!}
\lambda^{k-j} J_{n+1+j}(\lambda),
\end{eqnarray}
with no residual integral [notice that when Eq.\ (\ref{eqB-40}) is applied
with the upper index of $I^k_n$ greater by $1$ than its lower index,
then the second term on the RHS of this recursive formula
vanishes]. This leads, when applying the tilde-limit, to
\begin{equation}\label{eqB-110}
\widetilde{I}^k_n=0, \quad\text{for odd $k-n>0$}.
\end{equation}

The above results, Eqs.\ (\ref{eqB-30}), (\ref{eqB-70}), (\ref{eqB-90}),
and (\ref{eqB-110}), are summarized by Eq.\ (\ref{eq7-5}) in Sec.\
\ref{secVII}.

We further need now to calculate the integral given in Eq.\ (\ref{eq7-6}).
Integrating by parts, we express the primitive function of the
integrand, $\lambda^k J_n(\lambda) \ln\lambda$, for odd $k-n>0$, as
\begin{equation}\label{eqB-115}
I_n^{k{\rm (log)}}\equiv
\int \lambda^k J_n(\lambda) \ln\lambda\, d\lambda=
I^k_n(\lambda) \ln\lambda-\int \left[I^k_n(\lambda)/\lambda\right]
d\lambda.
\end{equation}
By virtue of Eq.\ (\ref{eqB-100}), the surface term on the RHS
here is dominated at small $\lambda$ by
$\propto \lambda^{(k+n+1)/2}J_{(k+n+1)/2}(\lambda)\ln\lambda\propto
\lambda^{k+n+1}\ln\lambda$, and therefore it vanishes at the limit
$\lambda\to 0$. However, this surface term diverges at the tilde-limit
$\lambda\to\infty$ (as $\lambda^{k-1/2} \ln\lambda$ times
oscillations with respect to $\lambda$).
This can be avoided by slightly modifying the definition
of a function's tilde-limit, by allowing the quantities
$B^{(j)}_l$ in Eq.\ (\ref{eqA-60}) to also admit the form
$B_l^{(j)}=a_j l^{b_j}\ln l\,\cos(\alpha_j l+\beta_j)$
(with $\alpha_j\ne 0$ for all $j$).
It can be shown that all features and results discussed in Appendix
\ref{AppA} concerning the tilde-limit remain valid also under
this wider definition.
With the revised definition of the tilde-limit, the surface term
in Eq.\ (\ref{eqB-115}) vanishes at the tilde-limit
$\lambda\to\infty$ as well as for $\lambda\to 0$.
Denoting $\tilde I_n^{k{\rm (log)}}\equiv \widetilde\lim_
{\lambda\to\infty}I_n^{k{\rm (log)}}$, we then have
\begin{eqnarray}\label{eqB-120}
\tilde I_n^{k{\rm (log)}}=
-\int_0^{\widetilde\infty}\left[I^k_n(\lambda)/\lambda\right] d\lambda=
-\sum_{j=0}^{(k-n-1)/2}(-1)^j\frac{(k-n-1)!!}{(k-n-1-2j)!!}\,
\widetilde{I}^{k-j-1}_{n+1+j},
\end{eqnarray}
where we have substituted for $I^k_n(\lambda)$ from Eq.\ (\ref{eqB-100}).
Provided that $k-n$ is an odd number, the difference between
the upper and lower indices of $\widetilde{I}^{k-j-1}_{n+1+j}$ is also
an odd number, $k-n-2-2j$. Therefore, by virtue of Eq.\ (\ref{eqB-110}),
we have $\widetilde{I}^{k-j-1}_{n+1+j}=0$ for any $j$ satisfying
$k-j-1>n+1+j$, i.e., $j<(k-n-2)/2$. We find that of all the terms summed
up in Eq.\ (\ref{eqB-120}), the only nonvanishing one is the one with
$j=(k-n-1)/2$. Hence,
\begin{eqnarray}\label{eqB-130}
\tilde I_n^{k{\rm (log)}}=
-(-1)^{(k-n-1)/2}(k-n-1)!!\,\widetilde{I}^{(k+n-1)/2}_{(k+n+1)/2}=
&&(-1)^{(k-n+1)/2}(k-n-1)!!(k+n-1)!!,
\end{eqnarray}
where, in the last equality, the value of
$\widetilde{I}^{(k+n-1)/2}_{(k+n+1)/2}$ has been inferred from Eq.\
(\ref{eqB-70}). This proves Eq.\ (\ref{eq7-6}).



\end{document}